\RequirePackage{commath, amsmath, amssymb, amsfonts}
\documentclass[10pt, a4paper]{iopart}

\usepackage{xcolor}
\usepackage{physics}
\usepackage{graphicx}
\usepackage{amsmath}
\usepackage{tikz}
\usepackage{hyperref}
\usepackage{physics}
\usepackage{amsmath}
\usepackage{mathdots}
\usepackage{yhmath}
\usepackage{cancel}
\usepackage{color}
\usepackage{siunitx}
\usepackage{array}
\usepackage{multirow}
\usepackage{amssymb}
\usepackage{gensymb}
\usepackage{tabularx}
\usepackage{booktabs}
\newcommand{\angstrom}{\textup{\AA}}
\newcommand{\lataone}{\ensuremath{\mathbf{a}_1}}
\newcommand{\latatwo}{\ensuremath{\mathbf{a}_2}}
\newcommand{\latbone}{\ensuremath{\mathbf{b}_1}}
\newcommand{\latbtwo}{\ensuremath{\mathbf{b}_2}}
\newcommand{\none}{\ensuremath{n_1}}
\newcommand{\ntwo}{\ensuremath{n_2}}
\newcommand{\mone}{\ensuremath{m_1}}
\newcommand{\mtwo}{\ensuremath{m_2}}
\newcommand{\rv}{\mathbf{r}}

\begin{document}

\tikzset{every picture/.style={line width=0.75pt}}
\title{Quantum point defects in 2D materials: The QPOD database}
\author{Fabian Bertoldo$^{\xi,*,\dagger}$, Sajid Ali$^{\xi,*,\dagger}$, Simone Manti$^{\xi,*}$, Kristian S. Thygesen$^\xi$}
\address{$^\xi$CAMD, Computational Atomic-Scale Materials Design, Department of Physics, Technical University of Denmark, 2800 Kgs. Lyngby Denmark}
\address{$^*$These authors contributed equally: Fabian Bertoldo, Sajid Ali, Simone Manti}
\address{$^{\dagger}$Corresponding authors: fafb@dtu.dk and sajal@dtu.dk}

\vspace{10pt}

\begin{abstract}
Atomically thin two-dimensional (2D) materials are ideal host systems for quantum defects as they offer easier characterisation, manipulation and read-out of defect states as compared to their bulk counterparts. Here we introduce the Quantum Point Defect (QPOD) database with more than 1900 defect systems comprising various charge states of 503 intrinsic point defects (vacancies and antisites) in 82 different 2D semiconductors and insulators. The Atomic Simulation Recipes (ASR) workflow framework was used to perform density functional theory (DFT) calculations of defect formation energies, charge transition levels, Fermi level positions, equilibrium defect and carrier concentrations, transition dipole moments, hyperfine coupling, and zero-field splitting. Excited states and photoluminescence spectra were calculated for selected high-spin defects. In this paper we describe the calculations and workflow behind the QPOD database, present an overview of its content, and discuss some general trends and correlations in the data. We analyse the degree of defect tolerance as well as intrinsic dopability of the host materials and identify promising defects for quantum technological applications. The database is freely available and can be browsed via a web-app interlinked with the Computational 2D Materials Database (C2DB). 

\end{abstract}

\maketitle
\ioptwocol

\section{Introduction}\label{sec:introduction}
Point defects are ubiquitous entities affecting the properties of any crystalline material. Under equilibrium conditions their concentration is given by the Boltzmann distribution, but strong deviations can occur in synthesised samples due to non-equilibrium growth conditions and significant energy barriers involved in the formation, transformation, or annihilation of defects. In many applications of semiconductor materials, in particular those relying on efficient carrier transport, the presence of defects has a detrimental impact on performance\cite{polman2016photovoltaic}. However, point defects in crystals can also be useful and form the basis for novel applications \textit{e.g.} in spintronics \cite{awschalom2013quantum,gomonay2018crystals}, quantum computing\cite{eckstein2013materials,gardas2018defects}, or quantum photonics\cite{lovchinsky2016nuclear,bradley2019ten,lovchinsky2016nuclear,bradley2019ten,sajid2020single,grosso2017tunable}. For such applications, defects may be introduced in a (semi)controlled manner \textit{e.g.} by electron/ion beam irradiation, implantation, plasma treatment or high-temperature annealing in the presence of different gasses\cite{fischer2021controlled}. \par
Over the past decade, atomically thin two-dimensional (2D) crystals have emerged as a promising class of materials with many attractive features including unique, easily tunable, and often superior physical properties\cite{ferrari2015science}. This holds in particular for their defect-based properties and related applications. Compared to point defects buried deep inside a bulk structure, defects in 2D materials are inherently surface-near making them easier to create, manipulate, and characterise\cite{lin2016defect}. Recently, single photon emission (SPE) has been observed from point defects in 2D materials such as hexagonal boron-nitride (hBN)\cite{tran2016quantum,grosso2017tunable,fischer2021controlled, sajid2020single}, MoS$_2$\cite{bertolazzi2017engineering}, and WSe$_2$\cite{chakraborty2015voltage,koperski2015single}, and in a few cases optically detected magnetic resonance (ODMR) has been demonstrated\cite{chejanovsky2021single,gottscholl2020initialization}. In the realm of catalysis, defects can act as active sites on otherwise chemically inert 2D materials \cite{ye2016defects,xie2020defect}.  
\par 
First-principles calculations based on density functional theory (DFT) can provide detailed insight into the physics and chemistry of point defects and how they influence materials properties at the atomic and electronic scales\cite{freysoldt2014first,janotti2007native,neugebauer1996gallium,dreyer2018first}. In particular, such calculations have become an indispensable tool for interpreting the results of experiments on defects, \textit{e.g.} (magneto)optical experiments, which in themselves only provide indirect information about the microscopic nature of the involved defect\cite{gupta2018two,fischer2021controlled,SajidAli2020Edgeeffects}.  In combination with recently developed tools for high-throughput workflow management\cite{jain2015fireworks,pizzi2016aiida,gjerding2021atomic,mortensen2020myqueue}, first-principles calculations have potential to play a more proactive role in the search for new defect systems with promising properties. Here, a major challenge is the notorious complexity of defect calculations (even when performed in low-throughput mode) that involves large supercells, local magnetic moments, electrostatic corrections, \textit{etc.} Performing such calculations for general defects in general host materials, requires a carefully designed workflow with optimised computational settings and a substantial amount of benchmarking\cite{sajid2018defect}. 
\par 
In this work, we present a systematic study of $503$ unique intrinsic point defects (vacancies and antisite defects) in $82$ insulating 2D host materials. The host materials were selected from the Computational 2D Materials Database (C2DB)\cite{haastrup2018computational,gjerding2021recent} after applying a series of filtering criteria. Our computational workflow incorporates the calculation and analysis of thermodynamic properties such as defect formation energies, charge transition levels (CTLs), equilibrium carrier concentrations and Fermi level position, as well as symmetry analysis of the defect atomic structures and wave functions, magnetic properties such as hyperfine coupling parameters and zero-field splittings, and optical transition dipoles. Defects with a high-spin ground state are particularly interesting for magneto-optical and quantum information technology applications. For such defects the excited state properties, zero phonon line energies, radiative lifetime, and photoluminescence (PL) lineshapes, were also calculated.  
\par 
The computational defect workflow was constructed with the Atomic Simulation Recipes (ASR)\cite{gjerding2021atomic} and executed using the MyQueue \cite{mortensen2020myqueue} task scheduler frontend. The ASR provides a simple and modular framework for developing Python workflow scripts, and its automatic caching system keeps track of job status and logs data provenance. Our ASR defect workflow adds to other ongoing efforts to automate the computational characterisation of point defects\cite{broberg2018pycdt,pean2017presentation,goyal2017computational,davidsson2021adaq}. However, to the best of our knowledge, the present work represents the first actual high-throughput study of point defects. All of the generated data is collected in the Quantum Point Defect (QPOD) database with over $1900$ rows and will be publicly available and accessible via a browsable web-service. The QPOD webpages are interlinked with the C2DB providing a seamless interface between the properties of the pristine host materials and their intrinsic point defects.   
\par 
The theoretical framework is based entirely on DFT with the Perdew Burke Ernzerhof (PBE) functional\cite{perdew1996generalized}. Charge transition levels are obtained using Slater-Janak transition state theory\cite{janak1978proof} while excited states are calculated with the DO-MOM method\cite{levi2020variational}. We note that PBE suffers from delocalisation errors\cite{mori2008localization}, which may introduce quantitative inaccuracies in the description of some localised defect states. While range-separated hybrid functionals represent the state-of-the art methodology for point defect calculations, such a description is currently too demanding for large-scale studies like the current. Moreover, thermodynamic properties of defects are generally well described by PBE\cite{lyons2017computationally}. 
\par
In Section \ref{sec:methodolgytheory} we describe the theory and methodology employed at the various computational steps of the workflow. Section \ref{sec:workflowinfrastructure} gives a general overview of the workflow, introduces the set of host materials and the considered point defects, and outlines the structure and content of the QPOD database web-interface. In Section \ref{sec:results} we present our main results. These include statistical overviews of host crystal and defect system properties, analysis of the effect of structural relaxations on defect formation energies and charge transition levels, an evaluation of the intrinsic (equilibrium) doping level in $58$ host materials and identification of a small subset of the host materials that are particularly defect tolerant. We also identify a few defect systems with promising properties for spin qubit applications or nanoscale magneto-optical sensing. Section \ref{sec:summaryoutlook} summarises the work and looks ahead.

\section{Theory and methodology}\label{sec:methodolgytheory}

\subsection{Supercell and defect structures}\label{subsec:supercell}
\begin{figure}[t]
    \centering
    \includegraphics[width=1\linewidth]{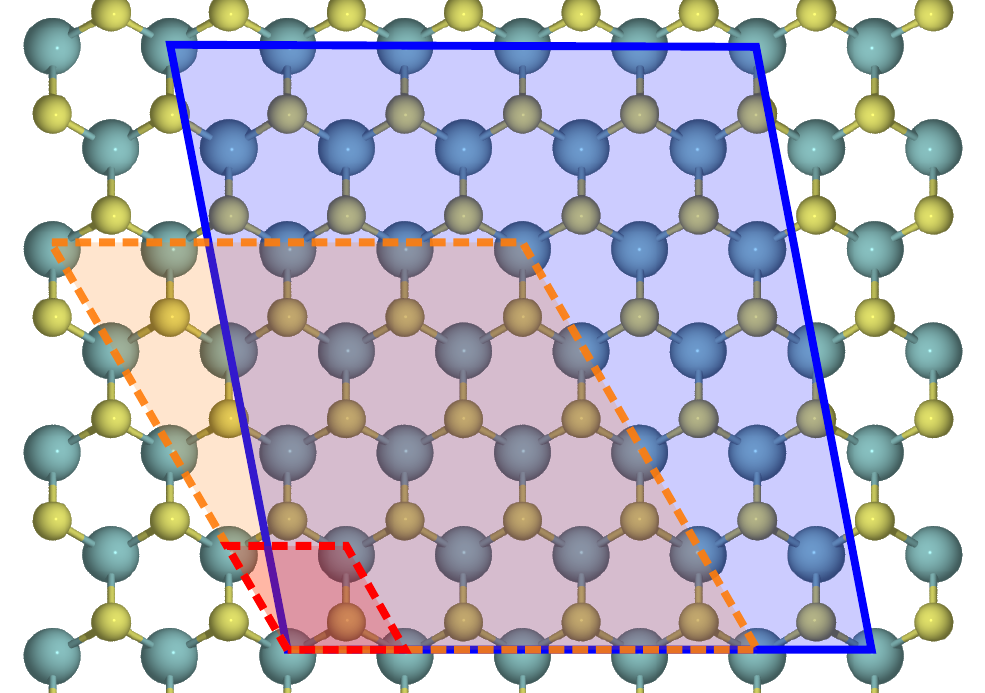}
    \caption{\textbf{Creation of defect supercells at the example of MoS$_2$.} Top view of a primitive unit cell of MoS$_2$ (red, dashed), an example of a conventional $4\times 4\times 1$ supercell (orange, dashed), as well as our different approach of a symmetry broken supercell (blue, solid).}
    \label{fig:2_2.1_supercell}
\end{figure}
A supercell in 2D can be created using linear combinations of the primitive unit cell vectors (\lataone, \latatwo). The corresponding supercell lattice vectors (\latbone, \latbtwo) are written as:
\begin{equation}\label{eq:b1}
    \latbone = \none\lataone + \ntwo\latatwo,
\end{equation}
\begin{equation}\label{eq:b2}
    \latbtwo = \mone\lataone + \mtwo\latatwo,
\end{equation}
where $\none$, $\ntwo$, $\mone$, $\mtwo$ are integers.\par
In this study, we apply an algorithm that finds the most suitable supercell according to the following criteria: (i) Set $\ntwo=0$ and create all supercells defined by $\none$, $\mone$, $\mtwo$ between 0 and 10. (ii) Discard combinations where $\mone = 0$ and $\none = \mtwo$. (iii) Keep only cells where the minimum distance between periodically repeated defects is larger than $15$ \AA. (iv) Keep supercells containing the smallest number of atoms. (v) Pick the supercell that yields the most homogeneous distribution of defects. We note that step (ii) is conducted in order to break the symmetry of the initial Bravais lattice, and step (iv) minimizes computational cost for a viable high-throughput execution.\par
Defects are introduced by analysing Wyckoff positions of the atoms within the primitive structure. For each non-equivalent position in the structure a vacancy defect and substitutional defects (called antisites in the following) are created (the latter by replacing a specific atom with another atom of a different species intrinsic to the host material). For the example of MoS$_2$ the procedure yields the following point defects: sulfur vacancy V\textsubscript{S}, molybdenum vacancy V\textsubscript{Mo}, as well as two substitutional defects Mo\textsubscript{S} and S\textsubscript{Mo} where Mo replaces the S atom and \textit{vice versa}. Each defect supercell created by this approach undergoes the workflow which is presented in Sec. \ref{sec:workflowinfrastructure}.

\subsection{Defect formation energy}
The formation energy of a defect X in charge state $q$ is defined by\cite{van1993first,zhang1991chemical}:
\begin{equation}\label{eq:eform}
E^{f}\left[\mathrm{X}^q\right] = E_{\mathrm{tot}}\left[\mathrm{X}^q\right] - E_{\mathrm{tot}}\left[\mathrm{bulk}\right] - \sum_i n_{i}\mu_i + qE_\mathrm{F}
\end{equation}
where $\mu_i$ is the chemical potential of the atom species $i$ and $n_i$ is the number of such atoms that have been added ($n_i > 0$) or removed ($n_i < 0$) in order to create the defect. In this work we set $\mu_i$ equal to the total energy of the the standard state of element $i$. We note, that in general $\mu_i$ can be varied in order to represent  $i$-rich and $i$-poor conditions\cite{van2004first}. For finite charge states, the defect formation energy becomes a function of the Fermi energy, $E_\mathrm{F}$, which represents the chemical potential of electrons. In equilibrium, the concentration of a specific defect type is determined by its formation energy, which in turn depends on $E_\mathrm{F}$. Imposing global charge neutrality leads to a self-consistency problem for $E_\mathrm{F}$, which we discuss in Sec. \ref{subsec:charge_neutrality}.
\par 
In general, the lower the formation energy of a particular defect is, the higher is the probability for it to be present in the material. In equilibrium, the defect concentration is given by the Boltzmann distribution,
\begin{equation}\label{eq:conc}
    C^{\mathrm{eq}}[\mathrm{X}^q] = N_\mathrm{X} g_{\mathrm{X}^q} \mathrm{exp}\left(-E^f\left[\mathrm{X}^q\right] / (k_\mathrm{B} T) \right),
\end{equation}
where $N_\mathrm{X}$ and $g_{\mathrm{X}^q}$ specifies the site and defect state degeneracy, respectively, $k_\mathrm{B}$ is Boltzmann's constant and $T$ is the temperature.
\par
\begin{figure}[t]
    \centering
    \includegraphics[width=1\linewidth]{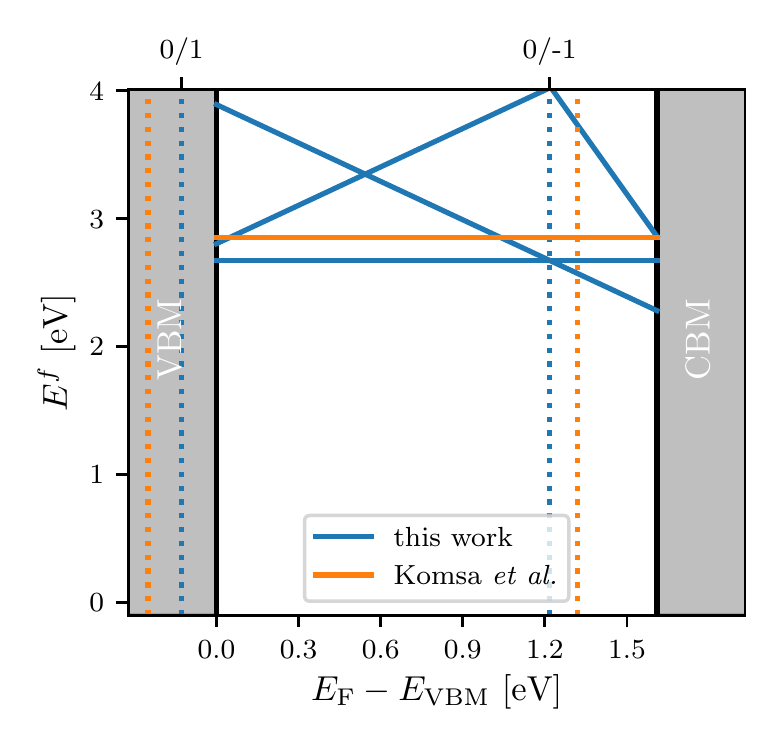}
    \caption{\textbf{Formation energy of V\textsubscript{S} in MoS$_2$ referenced to the standard states as a function of Fermi energy.} Calculated formation energy (blue, solid) as a function of the Fermi energy plotted together with the CTL (blue, dotted). The orange solid line (orange dotted lines) highlight the PBE-D calculated neutral formation energy (CTL of (0/-1) and (0/1)) of V\textsubscript{S} in MoS$_2$ taken from Komsa \textit{et al}. \cite{komsa2015native}.}
    \label{fig:2_2.2_formation}
\end{figure}

 
As an example, Fig. \ref{fig:2_2.2_formation} shows the formation energy (blue solid lines) of a sulfur vacancy V\textsubscript{S} in MoS$_2$ as a function of the Fermi level position. 
It follows that this particular defect is most stable in its neutral charge state ($q=0$) for 
low to mid gap Fermi level positions. The transition from $q=0$ to $q'=-1$ occurs for $E_\mathrm{F}$ around $1.2$ eV. The formation energy of the neutral V\textsubscript{S} and its CTLs reported in Ref. \cite{komsa2015native} are in good agreement with our values. The differences are below $0.2$ eV and can be ascribed to the difference in the employed supercells (symmetric vs. symmetry-broken) and xc-functionals (PBE-D vs. PBE). Moreover, in this work the CTLs are obtained using the Slater-Janak transition state theory (see Sec. \ref{sec:SJ}) while Ref. \cite{komsa2015native} used total energy differences with electrostatic corrections.

\par 
\subsection{Slater-Janak transition state theory}\label{sec:SJ}
\begin{figure*}[t]
    \centering
    \includegraphics[width=1.\linewidth]{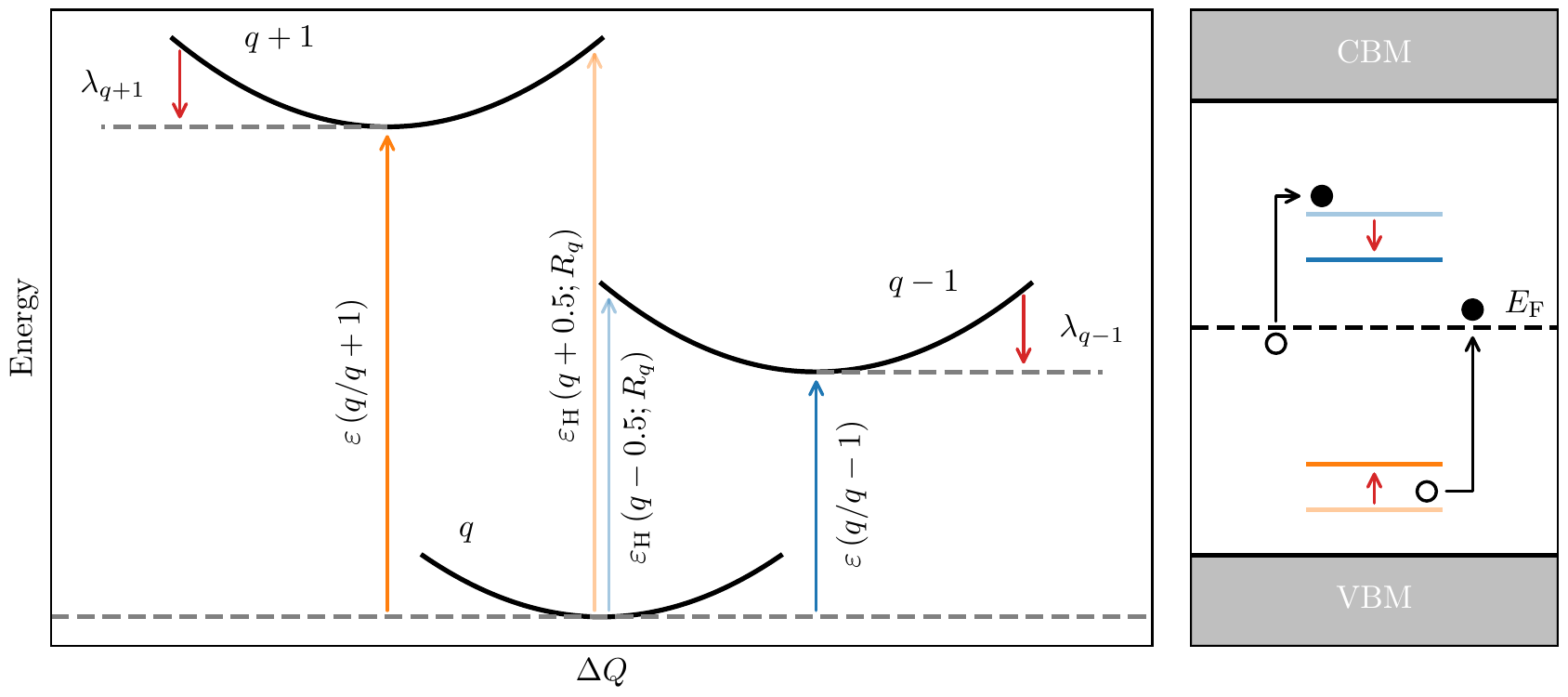}
    \caption{\textbf{Schematic view of the SJ approach to calculate charge transition levels.} Left: the optical charge transition levels are calculated starting from a given charge state $q$ by either adding (\textit{i.e.} $q-0.5$) or removing (\textit{i.e.} $q+0.5$) half an electron and calculating the HOMO energies at frozen atomic configuration (\textit{i.e.} $\varepsilon _\mathrm{H}\left(q\pm0.5;R_q\right)$, light colored arrows in left panel). Due to ionic relaxations one has to correct with the reorganization energies $\lambda _{q\pm1}$ to reduce the energy cost for the transition wrt. $E_\mathrm{F}$. Right: the reorganization energy for adding (removing) an electron has to be added (subtracted) from the optical transition levels (see red arrows in the right panel) in order to compute the thermodynamic CTLs $\varepsilon\left(q/q\pm1\right)$.}
    \label{fig:2_2.3_sj}
\end{figure*}
The prediction of charge transition levels requires the total energy of the defect in a different charge state. In the standard approach, the extra electrons/holes are included in the self-consistent DFT calculation and a background charge distribution is added to make the supercell overall charge neutral. In a post process step, the spurious interactions between periodically repeated images is removed from the total energy using an electrostatic correction scheme that involves a Gaussian approximation to the localised charge distribution and a model for the dielectric function of the material\cite{freysoldt2014first}. While this approach is fairly straightforward and unambiguous for bulk materials, it becomes significantly more challenging for 2D materials due to the spatial confinement and non-local nature of the dielectric function and the dependence on detailed shape of the neutralising background charge\cite{komsa2014charged,liu2020extrapolated,xia2020evaluation,xiao2020realistic}. 

To avoid the difficulties associated with electrostatic corrections, we rely on the Slater-Janak (SJ) theorem\cite{janak1978proof}, which relates the Kohn-Sham eigenvalue $\varepsilon_i$ to the derivative of the total energy with respect to the orbital occupation number 
 $n_i$,
\begin{equation}    
\frac{\partial E}{\partial n_i} = \varepsilon_i(n_i).
\end{equation}
The theorem can be used to express the difference in ground state energy between two charge states as an integral over the eigenvalue as its occupation number is changed from 0 to 1. This approached, termed SJ transition state theory, has been used successfully used to evaluate core-level shifts in random alloys\cite{goransson2005numerical}, and CTLs of impurities in GaN\cite{sanna2008validity}, native defects in LiNbO$_3$\cite{li2014modeling}, and chalcogen vacancies in monolayer TMDs  \cite{pandey2016defect}. Assuming a linear dependence of $\varepsilon_i$ on $n_i$ (which holds exactly for the true Kohn-Sham system), the transition energy between two localised states of charge $q$ and $q'=q\pm 1$ can be written as
\begin{equation}
    \varepsilon\left(q/q'\right) =  \left\{
    \begin{array}{ll}
         \varepsilon_\mathrm{H}\left(q + \frac{1}{2}; R_q\right)-\lambda_{q'}, & q' = q + 1 \\
         & \\
        \varepsilon_\mathrm{H}\left(q - \frac{1}{2}; R_q\right)+\lambda_{q'}, & q' = q - 1.
    \end{array}
    \right.
\end{equation}
Here, $\varepsilon_\mathrm{H}$ represents highest eigenvalue with non-zero occupation, \textit{i.e.} the half occupied state, and $R_q$ refers to the configuration of charge state $q$. The reorganisation energy is obtained as a total energy difference between equal charge states
\begin{equation}
    \lambda_{q'} = E_{\mathrm{tot}}\left(q'; R_{q'}\right) - E_{\mathrm{tot}}\left(q'; R_{q}\right).
\end{equation}
Note that the reorganisation energy is always negative. The relevant quantities are illustrated graphically in Figure \ref{fig:2_2.3_sj}. 

It has been shown for defects in bulk materials that the CTLs obtained from SJ theory are in good agreement with results obtained from total energy differences\cite{goransson2005numerical,li2014modeling}. For 2D materials, a major advantage of the SJ method is that it circumvents the issues related to the electrostatic correction, because it completely avoids the comparison of energies between supercells with different number of electrons. The Kohn-Sham eigenvalues of a neutral or (partially) charged defect supercell are referenced relative to the electrostatic potential averaged over the PAW sphere of an atom located as far as possible away from the defect site (typically around $7$ {\AA} depending on the exact size of the supercell). By performing the potential average over an equivalent atom of the pristine 2D layer, we can reference the Kohn-Sham eigenvalue of the defect supercell to the VBM of the pristine material. As an alternative to averaging the potential around an atom, the asymptotic vacuum potential can be used as reference. We have checked that the two procedures yield identical results (usually within 0.1 eV), but prefer the atom-averaging scheme as it can be applied to bulk materials as well.

\subsection{Equilibrium defect concentrations} \label{subsec:charge_neutrality}
\begin{figure*}[t]
    \centering
    \includegraphics[width=1\linewidth]{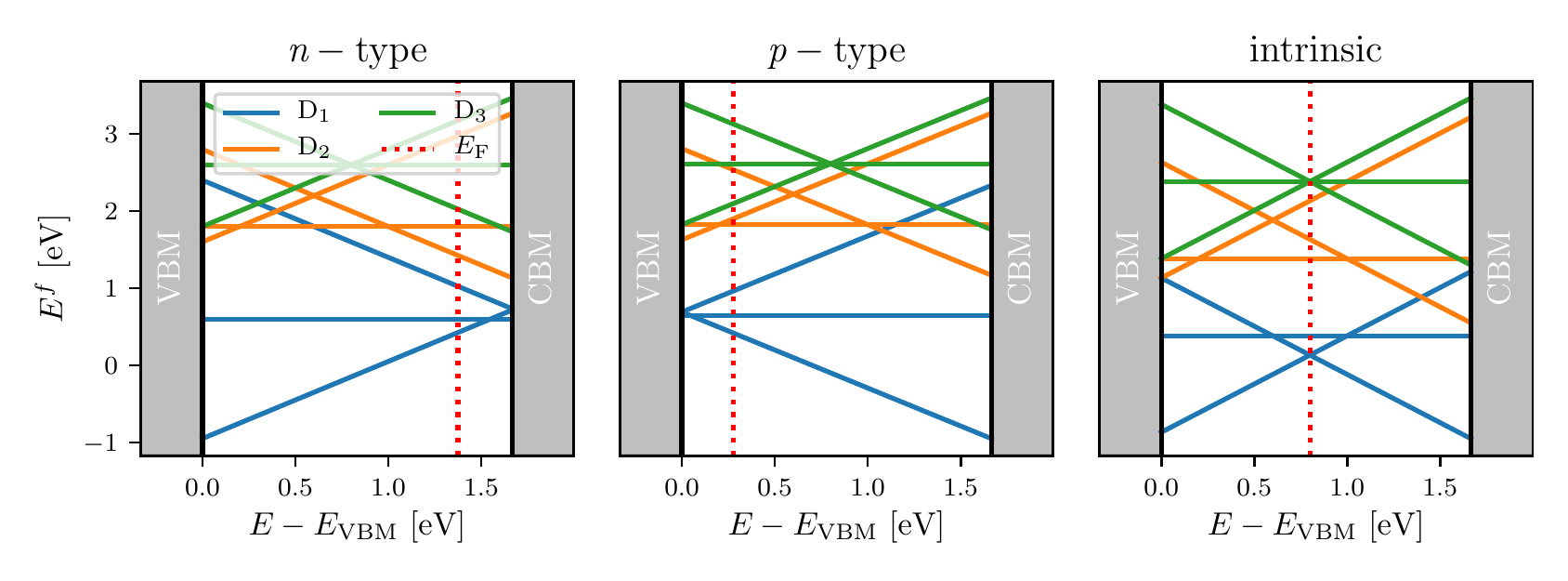}
    \caption{\textbf{Intrinsic dopability of defect systems.} Formation energies with respect to the standard states as a function of energy for three mock-sytems with three defect types present, respectively. Left: \textit{n}-type dopable regime ($E_\mathrm{F}$ close to CBM) with dominating donor defect. Middle: \textit{p}-type dopable regime ($E_\mathrm{F}$ close to VBM) with dominating acceptor defect. Right: intrinsic material (neither \textit{p}- nor \textit{n}-dopable due to the presence of competing acceptor and donor defects.)}
    \label{fig:dopability}
\end{figure*}
According to Eqs. (\ref{eq:eform}) and (\ref{eq:conc}), the formation energy of charged defects, and therefore their equilibrium concentration, is a function of the Fermi level. The Fermi level position of the system in thermal equilibrium is then determined self-consistently from a requirement of charge neutrality\cite{buckeridge2019equilibrium}
\begin{equation}\label{eq:charge_neutrality}
     \sum_X\sum_q qC[\mathrm{X}^q] = n_0-p_0
\end{equation}
where the sum is over structural defects $\mathrm{X}$ in charge states $q$, and $n_0$ and $p_0$ are the electron and hole carrier concentrations, respectively. The latter are given by
\begin{equation}
    n_0 = \int_{E_{\mathrm{gap}}}^\infty dE f(E)\rho(E),
\end{equation}
\begin{equation}
    p_0 = \int_{-\infty}^0 dE [1-f(E)]\rho(E)
\end{equation}
where $\rho(E)$ is the local density of states, $f(E)=1/\exp[(E-E_\mathrm{F})/k_\mathrm{B}T]$ is the Fermi-Dirac distribution, and the energy scale is referenced to the valence band maximum of the pristine crystal. 
\par 
Under the assumption that all the relevant defects, \textit{i.e.} the intrinsic defects with the lowest formation energies, are accounted for, Eq. (\ref{eq:charge_neutrality}) will determine the Fermi level position of the material in thermal equilibrium. The equilibrium Fermi level position determines whether a material is intrinsically \textit{p}-doped, \textit{n}-doped, or intrinsic. The three different cases are illustrated schematically in Fig. \ref{fig:dopability}. For the \textit{n}-type case (left panel), the most stable defect is D\textsubscript{1} in charge state $+1$. Thus, charge carriers are transferred from the defect into the conduction band resulting in a Fermi level just below the CBM. Similarly, for the \textit{p}-type case (middle panel) the defect D\textsubscript{1} in charge state $-1$ is the most stable. Consequently, charge carriers are promoted from the valence band into the defect resulting in $E_\mathrm{F}$ close to the VBM. In the right panel, donor and acceptor states are competing, which results in an effective cancelling of the \textit{p}- and \textit{n}-type behavior, pinning the Fermi level in the middle of the band gap. In Section \ref{sec:dopability} we analyse the intrinsic carrier types and concentrations of $58$ host materials where we set $g_{\mathrm{X}^q}$ and $N_\mathrm{X}$ in Eq. (\ref{eq:conc}) to one.  
\subsection{Symmetry analysis}\label{subsec:symmetry}
All states with an energy in the band gap are classified according to the symmetry group of the defect using a generalization of the methodology previously implemented in GPAW for molecules \cite{kaappa2018point}. In a first step, the point group, $G$, of the defect is determined. To determine $G$ we first reintroduce the relaxed defect into a supercell that preserves the symmetry of the host material; precisely, a supercell with basis vectors defined by setting $n_1=m_2=n$ and $n_2=m_1=0$ in Eqs. (\ref{eq:b1}, \ref{eq:b2}). We then use spglib\cite{spglib2009togo} to obtain $G$ as the point group of the new supercell. We stress that the high-symmetry supercell is only used to determine $G$ while all actual calculations are performed for the low-symmetry supercell as described in Sec. \ref{subsec:symmetry}.  

The defect states in the band gap are labeled according to the irreducible representations (irrep) of $G$. To obtain the irrep of a given eigenstate, $\psi_n(\rv)$, we form the matrix elements 
\begin{equation}\label{eq:gamma}
\Gamma(\mathrm{R}) = \Gamma_{nn}(\mathrm{R}) = \int d\rv\; \psi_n(\rv)^* \mathrm{R}\, \psi_n(\rv)
\end{equation}
where $\mathrm{R}$ is any symmetry transformation of $G$. It follows from the orthogonality theorem\cite{cornwell1997group} that the vector $\Gamma(\mathrm{R})$ can be expanded in the character vectors, $\chi^{(\alpha)}(\mathrm{R})$,
\begin{equation}
\Gamma(\mathrm{R}) = \sum_\alpha\, c_\alpha\, \chi^{(\alpha)}(\mathrm{R})
\label{eqn:overlapexp}
\end{equation}
where the quantity $c_\alpha$ represents the fraction of $\psi_n$ that transforms according to the irrep $\alpha$. For any well localised defect state, all the $c_\alpha$ will be zero except one, which is the irrep of the state. In general, less localised states will not transform according to an irrep of $G$. That is because the wave functions are calculated in the symmetry-broken supercell, see Sec. \ref{subsec:supercell}, and therefore will have a lower symmetry than the defect. To exclude such low-symmetry tails on the wave functions, the integral in Eq. (\ref{eq:gamma}) is truncated beyond a cut-off radius measured from the center of symmetry of the defect.
\par
\begin{figure*}[t]
    \centering
    \includegraphics[width=1\linewidth]{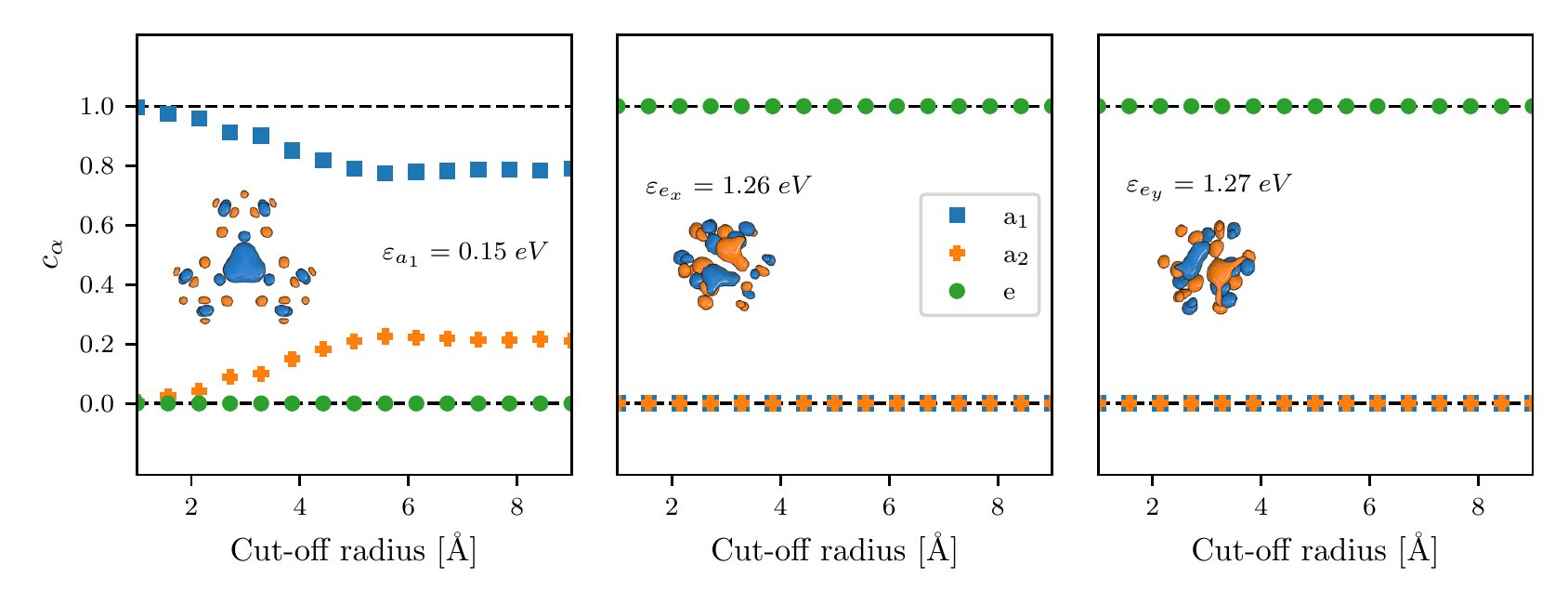}
    \caption{\textbf{Orbital symmetry labels for different cutoff radii.} All defect states with an energy inside the band gap of the host material are classified according to the irreps of the point group of the defect. The $c_\alpha$ coefficient is a measure of the degree to which a defect state transforms according to irrep $\alpha$. Performing the symmetry analysis within a radius of a few $\angstrom$ of the center of symmetry leads to a correct classification of the totally symmetric a$_1$ state and the degenerate e$_x$ and e$_y$ states of the neutral sulfur vacancy in MoS\textsubscript{2}. The isosurfaces of the orbitals are shown as insets and the energy eigenvalues reference to the VBM.}
    \label{fig:2_2.5_symmetry}
\end{figure*}

As an example, Figure \ref{fig:2_2.5_symmetry} shows the coefficients $c_\alpha$ for the in-gap states of the (neutral) sulfur vacancy in MoS\textsubscript{2}. This is a well studied prototypical defect with C$_{3v}$ symmetry. The absence of the chalcogen atom introduces three defect states in the gap: a totally symmetric $a_1$ state close to the VBM and two doubly degenerate mid-gap states $(e_x,e_y)$. The symmetry coefficients $c_\alpha$ correctly captures the expected symmetry of the states\cite{komsa2015native}, at least for small cut-off radii. The effect of employing a symmetry-broken supercell can be seen on the totally symmetric state a$_1$, which starts to be mixed with the antisymmetric a$_2$ irrep as a function of the radius, while there is no effect on the degenerate $e_x$ state. Therefore a radius of 2 {\AA} is used in the database to catch the expected local symmetry of the defect.
\subsection{Transition dipole moment}
The transition dipole moment is calculated between all single particle Kohn-Sham states inside the band gap, 
\begin{equation}\label{eq:dipole}
\mu_{nm}= \bra {\psi_n} \mathbf {\hat {r}} \ket {\psi_m} = \frac {i\hbar}{m_e}\frac{\bra {\psi_{n}} \mathbf {\hat{p}} \ket {\psi_{m}}}{\varepsilon_{n}-\varepsilon_{m}},
\end{equation}
where $ \mathbf {\hat {r}} $ is the dipole operator, $\mathbf {\hat {p}}$ is the momentum operator, and $m_e$ is the electron mass.  The transition dipole moment yields information on the possible polarization directions and oscillator strength of a given transition. In this work, the transition dipoles between localised defect states are calculated in real space, \textit{i.e.} using the first expression in Eq. (\ref{eq:dipole}), after translating the defect to the center of the supercell.
\subsection{Radiative recombination rate and life time}
Radiative recombination refers to the spontaneous decay of an electron from an initial high energy state to a state of lower energy upon emitting a photon. The rate of a spin-preserving radiative transition between an initial state $\psi_m$ and a final state $\psi_n$ is given by \cite{stoneham2001theory}
\begin{equation}
    \Gamma^{\mathrm{rad}}_{nm} = 1/\tau_{\mathrm{rad}} = \frac{ E^3_{\mathrm{ZPL}} \mu_{nm}^2} {3\pi \epsilon_0c^3 \hbar^4}.
\end{equation}
Here, $\epsilon_0$ is the vacuum permitivity, $E_{\mathrm{ZPL}}$ is the zero phonon line (ZPL) energy of the transition (see Sec. \ref{subsec:excited_methodology}), and $\mu_{nm}$ is the transition dipole moment defined in Eq. (\ref{eq:dipole}). The ZPL energy includes the reorganisation energy due to structural differences between the initial and final states, which can be on the order of 1 eV. Consequently, an accurate estimate of the radiative lifetime requires a geometry optimisation in the excited state. Since this step is not part of our general workflow, but is only performed for selected defects, radiative lifetimes are currently only available for a few transitions.
\subsection{Hyperfine coupling}
Hyperfine (HF) coupling refers to the interaction between the magnetic dipole associated with a nuclear spin, $\mathbf {\hat {I}}^N$, and the magnetic dipole of the electron-spin distribution, $\mathbf{\hat {S}}$. For a fixed atomic nuclei, $N$, the interaction is written
\begin{equation}
\hat H_{\mathrm {HF}}^N = \sum_{i,j} \hat {S}_i A_{ij}^{N} \hat {I}^{N}_j
\end{equation}
where the hyperfine tensor $\mathbf{A}^N$ is given by 
\begin{equation}
 \begin{aligned}
 A^N_{ij} &= \frac{2 \alpha^2 g_e m_e}{3 M_N}
    \int \delta_T(\mathbf{r}) \rho_s(\mathbf{r}) d\mathbf{r}\\
        &  + \frac{\alpha^2 g_e m_e}{4 \pi M_N} \int \frac{3 r_i r_j - \delta_{ij} r^2}{r^5}
    \rho_s(\mathbf{r}) d\mathbf{r}.\end{aligned}
\end{equation}
The first term represents the isotropic Fermi-contact term, which results from a non-vanishing electron spin density $\rho_s(\mathbf{r})$ at the centre of the nucleus. $\delta_T(\mathbf{r})$ is a smeared out $\delta$-function which appears in place of an ordinary $\delta$-function in the non-relativistic formulation of the Fermi-contact term and regulates the divergence of the \textit{s}-electron wave function at the atomic core\cite{blochl2000first}. $\alpha$ is the fine structure constant, $m_e$ is the electron mass, $M_N$ is the mass of atom $N$, and $g_e$ is the gyromagnetic ratio for electrons. The second term represents the anisotropic part of the hyperfine coupling tensor and results from dipole-dipole interactions between nuclear and electronic magnetic moments. 

\begin{figure}[t]
    \centering
    \includegraphics[width=1\linewidth]{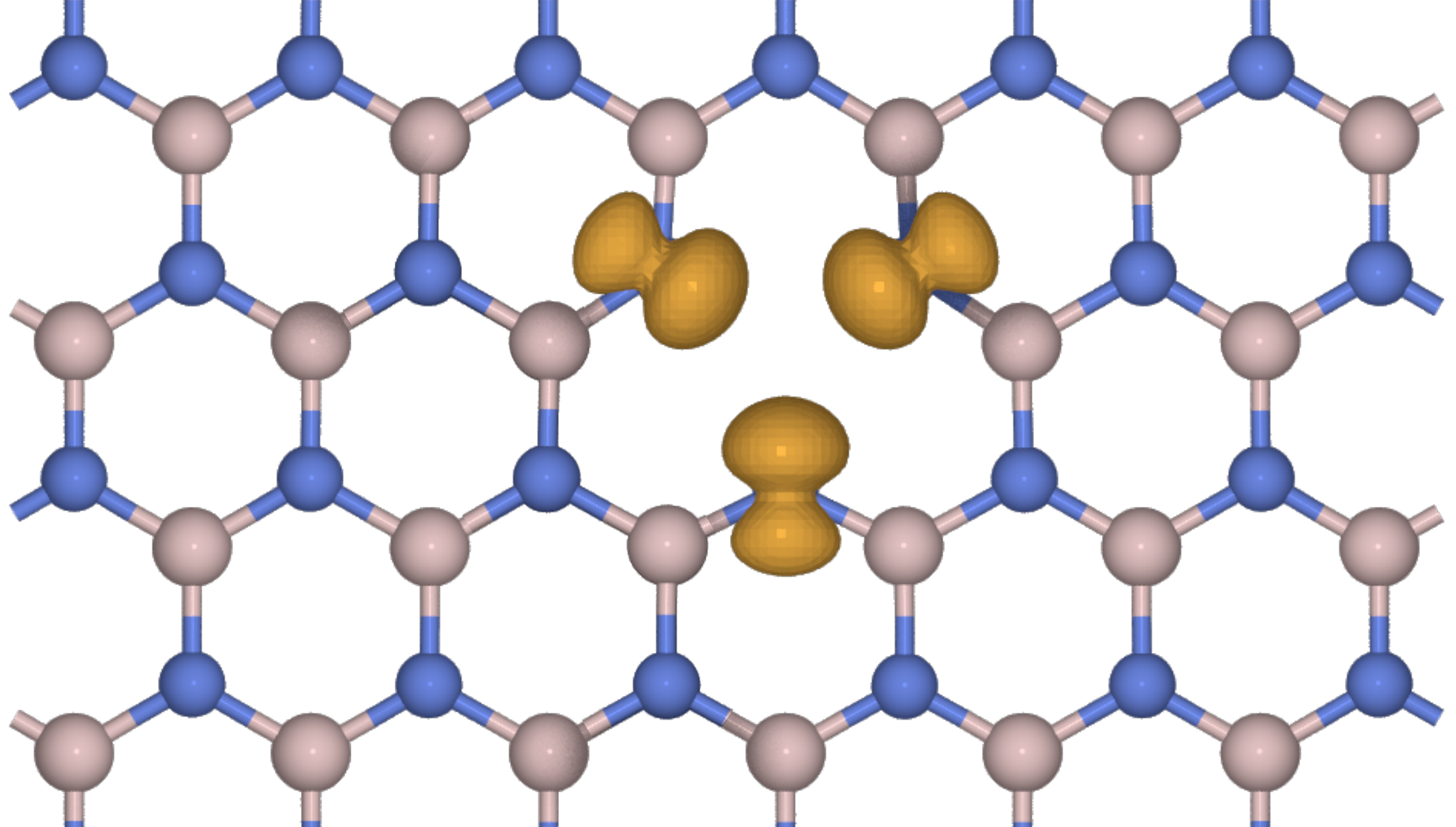}
    \caption{Iso-surface of the calculated spin-density (isovalue of 0.002 ${\left| e\right|}{\mathrm{\AA}}^{-3}$) for the V$_\mathrm{B}^{-}$ defect in hexagonal boron nitride.}
    \label{fig:2_2.9_isosurface}
\end{figure}

The Fermi-contact term $a = \mathrm{Tr}\left(\mathbf{A}^N\right) / 3$ depends only on the $l=0$ component of the spin density at the nucleus. In contrast, the anisotropic term $A_{ij}^N - a$ is sensitive to the $l>0$ components of the spin density near the nucleus. The hyperfine tensor thus provides direct insight into the electron spin distribution near the corresponding nucleus, and a direct comparison of the calculated HF coupling constants with electron paramagnetic resonance spectroscopy measurements can help to identify the nature of defect centers\cite{sajid2018defect, SajidAli2020Edgeeffects}. As an illustration, Figure \ref{fig:2_2.9_isosurface} shows the iso-surface of the spin density of the V$_\mathrm{B}^-$ defect in hexagonal boron nitride.   

In the present work, the $\mathbf A^N$ tensors are calculated for all atoms of the supercell, and its eigenvalues, also known as the HF principal values, are reported in the QPOD database. 

\subsection{Zero field splitting}
Zero field splitting (ZFS) refers to the splitting of magnetic sub levels of a triplet defect state due to the magnetic dipole-dipole interaction of the two electron spins that takes place even in the absence of an external magnetic field.  A triplet ($S = 1$) defect state can be described by a spin Hamiltonian of the following form
\begin{equation}\label{eq:ZFS}
\hat    H_{\mathrm{ZFS}} = \sum_{ij}\hat S_iD_{ij}\hat S_j, 
\end{equation}
where $\mathbf{\hat S}$ is the total spin operator and $\mathbf{D}$ is the ZFS tensor given by 
\begin{equation}
 \begin{aligned}
 D_{ij} &= \frac{\alpha^2 g_e m_e}{4 \pi} \int |\phi_{ij}(\mathbf {r}_{1},\mathbf {r}_{2})|^2 \frac{3 r_{i} r_{j} - \delta_{ij}{r}^2}{r^5} d\mathbf{r}_1\mathbf{r}_2.
    \end{aligned}
\end{equation}      
Eq. (\ref{eq:ZFS}) can also be written as
\begin{eqnarray}
\hat H_\mathrm{ZFS} &= D_{xx} \hat S_x^2+D_{yy} \hat S_y^2+D_{zz} \hat S_z^2 \nonumber \\     
        &= D(\hat S_z^2-S( S+1)/3)+E(\hat S_x^2-\hat S_y^2 ),
        \end{eqnarray}
where $D=3D_{zz}/2$ and $E=(D_{xx}-D_{yy})/2$ are called axial and rhombic ZFS parameters, respectively.  $D$ describes the splitting between the $m_s=\pm 1$ and $m_s=0$ magnetic sub-levels, while $E$ describes the splitting of the $m_s=\pm1$ sub-levels.  $D$ is generally zero for a spherically symmetric wave function because there is no direction in which electrons of a triplet can move to minimize the repulsive dipole-dipole interaction. However, for non-spherical wave functions, $D$ will be non-zero and lift the degeneracy of magnetic sub-levels $m_s=±1$ and $m_s=0$. A positive value of $D$ will result from an oblate spin-distribution, while a negative value will result from a prolate spin-distribution. The value of $E$ will be zero for axially symmetric wave functions. 

\subsection{Excited states}\label{subsec:excited_methodology}
 In Kohn-Sham DFT, excited electronic states can be found by solving the Kohn-Sham equations with non-Aufbau occupations of the orbitals. Often this approach is referred to as the delta self-consistent field (Delta-SCF) method\cite{gunnarsson1976exchange}. 
 \begin{figure}[t]
    \centering
    \includegraphics[width=1\linewidth]{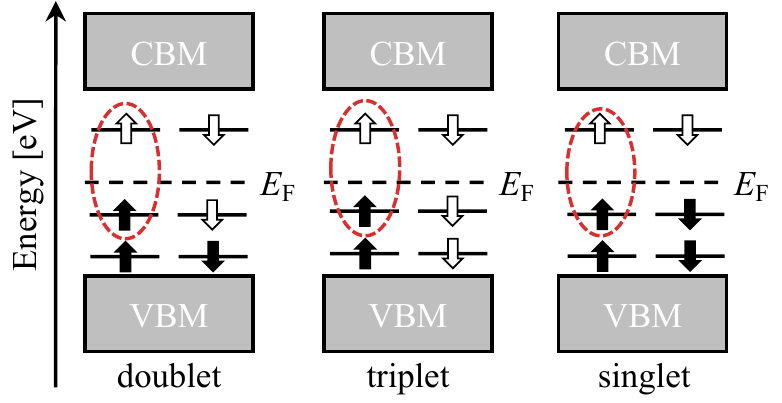}
    \caption{Possible occupancies resulting in different spin configurations of the defect systems for the excited state calculations. The states involved in the excitation are encircled red.}
    \label{fig:2_2.12_triplet}
\end{figure}
The excited state solutions are saddle points of the Kohn-Sham energy functional. Unfortunately, the Delta-SCF approach often struggle to find such solutions, especially when nearly degenerate states are involved. The Delta-SCF method fails in particular for cases involving charge transfer or Rydberg states \cite{levi2020variational}. This is due to a significant rearrangement of charge density between orbitals with similar energy.

Therefore, in the present work we use an alternative to the conventional Delta-SCF method, namely a method based on the direct optimization (DO) of orbital rotations by means of an efficient quasi-Newton method for saddle points, in combination with the maximum overlap method (MOM)\cite{levi2020variational}. The MOM ensures that the character of the states is preserved during the optimization procedure. In the DO-MOM method, convergence towards the \textit{n}-th order saddle point is guided by an appropriate pre-conditioner based on an approximation to the Hessian with exactly $n$ negative eigenvalues. This method ensures fast and robust convergence of the excited states, as compared to conventional Delta-SCF \cite{levi2020variational} methods. 

The DO-MOM method has been previously used for the calculation of excited state spectra of molecules \cite{levi2020variational}. However, this work is the first application of the method to defect states. 
We have benchmarked the method for a range of point defects in diamond, hBN, SiC, and MoS$_2$, and established that the method yields results in good agreement with conventional Delta-SCF calculations.
 
The most frequently occurring point defect spin configurations are sketched in Fig. \ref{fig:2_2.12_triplet}. For the doublet and triplet spin configuration, both the ground and excited states can be expressed as single Slater determinants. However, for the singlet spin configuration, the ground state is a closed shell singlet, while the excited state, as a result of the single excitation in either spin channel, will result in an open shell singlet state, which cannot be expressed as a single Slater determinant. This open shell singlet state can, however, be written as a sum of two Slater determinants of the form 
$|a\uparrow,b\downarrow\rangle$ and $|a\downarrow,b\uparrow\rangle$, each of which represent a mixed singlet-triplet state accessible with Delta-SCF. This allows us to obtain the singlet energy as
\begin{equation}\label{eq:OSS}
E_{s} = 2E_{st} - E_{t}.
\end{equation}
Here, $E_{st}$ is the DFT energy calculated by setting the occupancy for the open-shell singlet state, while $E_s$ and $E_t$
are the energies of the corresponding singlet and triplet states. Note that the latter can be represented as a Slater determinant. 

\begin{figure}[t]
    \centering
    \includegraphics[width=1.\linewidth]{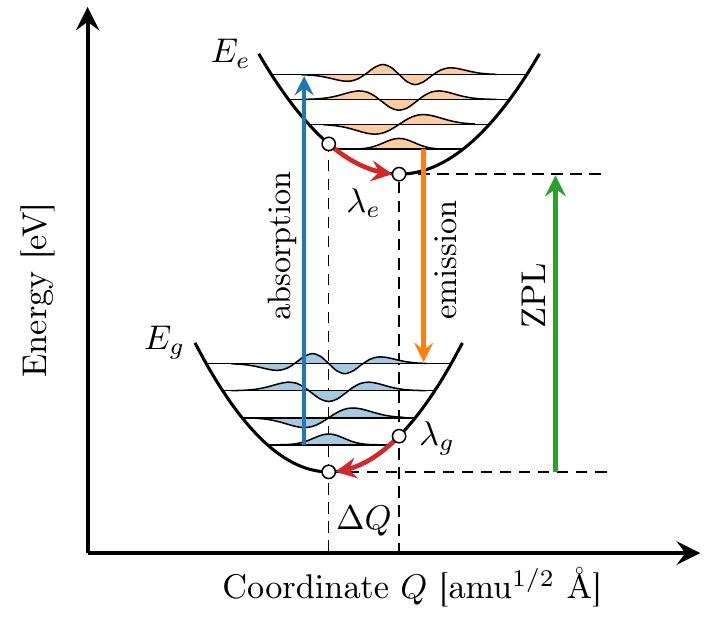}
    \caption{\textbf{Schematic CC diagram of a ground state to excited state transition.} Absorption and emission to and from the excited state (blue, orange) as well as their respective reorganization energies $\lambda_\mathrm{e,g}$ as a function of energy and configuration coordinate $Q$. The zero phonon line transition (ZPL) is visualized in green.}
    \label{fig:2_2.12_cc}
\end{figure}

Photoluminescence spectra of selected point defects were calculated using a generating function approach\cite{alkauskas2014first} outlined below. First, the mass weighted difference between atomic coordinates in the ground and excited electronic states is computed as follows
\begin{equation}
\Delta Q = \sqrt {\sum_\alpha m_\alpha \Delta R_\alpha^2},
\end{equation}
where the sum runs over all the atoms in the supercell. Afterwards, the partial Huang-Rhys factors are computed as 
\begin{equation}
S_k  = \frac{1}{2\hbar} \omega_k Q_k^2,
\end{equation}
where $Q_k$ is the projection of the lattice displacement on the normal coordinates of the ground state described by phonon mode $k$. The electron-phonon spectral function, which depends on the coupling between lattice
displacement and vibrational degrees of freedom, is then obtained by summing over all the modes $k$
\begin{equation}
S(\omega)  = \sum_k S_k \delta(\omega - \omega_k).
\end{equation}
The integral over the electron-phonon spectral function gives the (total) HR factor of the transition. The above electron-phonon spectral function is fed into a generating function to produce the photoluminescence lineshape\cite{alkauskas2014first}.

\section{QPOD workflow and database}\label{sec:workflowinfrastructure}
\subsection{The QPOD workflow}\label{subsec:workflow_QPOD}
\begin{figure*}
  \centering
  \includegraphics{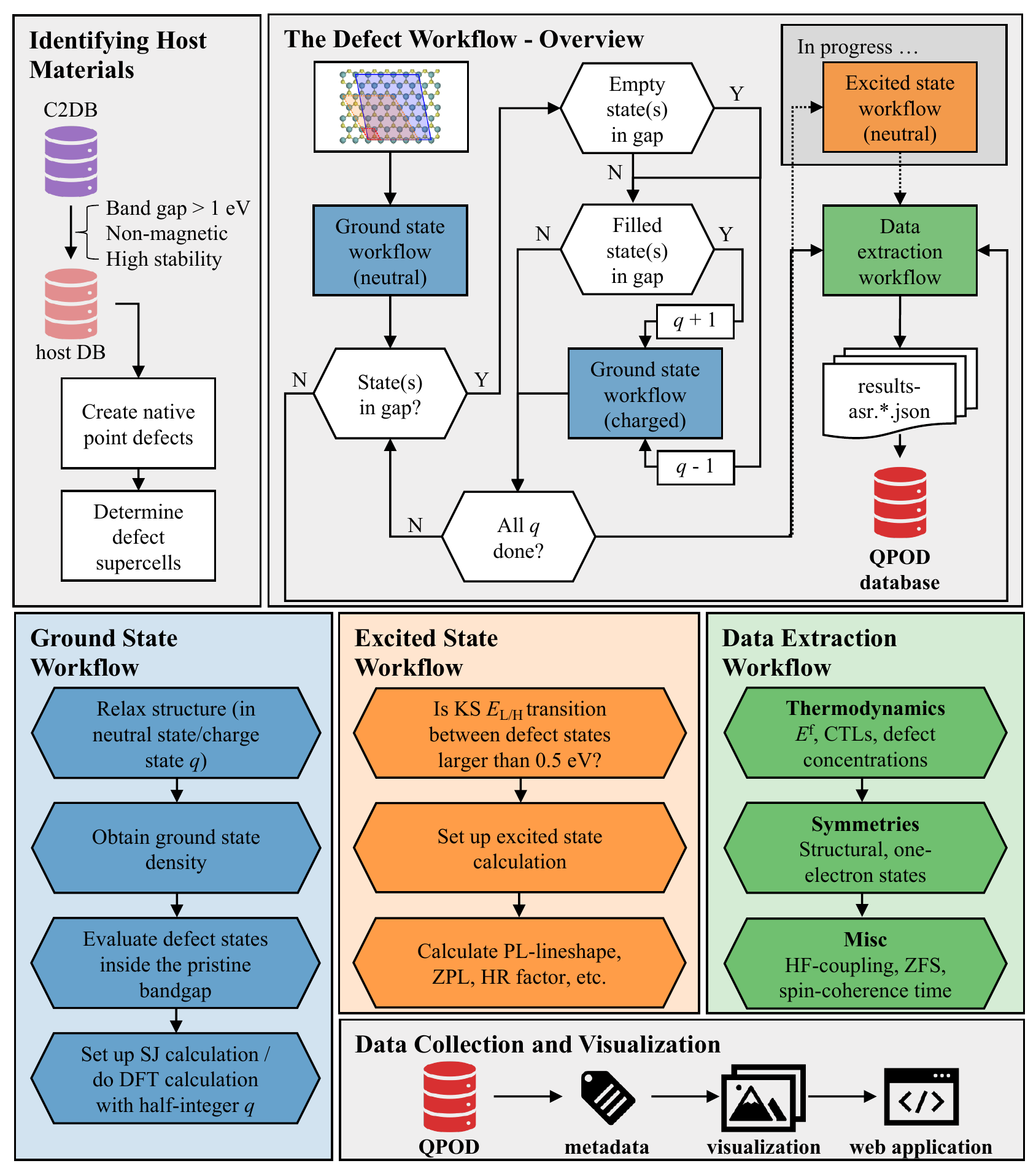}
  \caption{\textbf{The workflow behind the defect database.} First, starting from C2DB suitable host materials are identified. With the ASR recipe for defect generation the defect supercells are set up and enter the ground state workflow for neutral systems. Afterwards, depending on the nature of the defect states inside the gap, charged calculations are conducted within the charged ground state workflow. If all charged calculations for a specific system are done ($|q| < 3$), physical results are collected using the data extraction workflow and thereafter saved in the defect database. The database is equipped with all necessary metadata, and together with various visualization scripts the browseable web application is created. The excited state workflow is executed for selected systems as discussed in Sec. \ref{subsec:quantum_applications}.}
  \label{fig:3.1_workflow}
\end{figure*}
The backbone of the QPOD database is represented by a high-throughput framework based on the Atomic Simulation Recipes (ASR) \cite{gjerding2021atomic} in connection with the MyQueue \cite{mortensen2020myqueue} task and workflow scheduling system. Numerous recipes, designed particularly for the evaluation of defect properties, have been implemented in ASR and have been combined in a central MyQueue workflow to generate all data for the QPOD database.\par 
The underlying workflow is sketched in Fig. \ref{fig:3.1_workflow}, and will be described briefly in the following. As a preliminary step the C2DB \cite{haastrup2018computational,gjerding2021recent} is screened to obtain the set of host materials. Only non-magnetic, thermodynamically and dynamically stable materials with a PBE band gap of $E_{\mathrm{gap}}^{\mathrm{PBE}} > 1$ eV are selected as host materials. These criteria result in $281$ materials of which we select $82$ from a criterion of $N_e \times V_\mathrm{supercell} < 0.9$ {\AA}$^3$ combined with a few handpicked experimentally known and relevant 2D materials (MoS$_2$, hBN, WS$_2$, MoSe$_2$). It is important to mention, that some host materials exist in different phases (same chemical formula and stoichiometry, but different symmetry), \textit{e.g.} $2$H-MoS$_2$ and $1$T-MoS$_2$. In these cases we only keep the most stable one, \textit{i.e.} $2$H-MoS$_2$ for the previous example.\par 

For each host material, all inequivalent vacancies and antisite defects are created in a supercell as described in Sec. \ref{subsec:supercell}. Each defect enters the ground state workflow, which includes relaxation of the neutral defect structure, calculation of a well-converged ground state density, identification and extraction of defect states within the pristine band gap, and SJ calculations with half-integer charges $q$. If there are no states within the gap, the defect system directly undergoes the data extraction workflow and is stored in the QPOD database.
\par 
For systems with in-gap states above (below) $E_\mathrm{F}$, an electron is added (removed) and the charged structures are relaxed, their ground state calculated and the states within the band gap are examined again up to a maximum charge of $+3$/$-3$. Once all charge states have been relaxed and their ground state density has been evaluated, the data extraction workflow is executed. Here, general defect information (defect name, defect charge, nearest defect-defect distance, \textit{etc.}), charge transition levels and formation energies, the equilibrium self-consistent Fermi level, equilibrium defect concentrations, symmetries of the defect states within the gap, hyperfine coupling, transition dipole moments, \textit{etc.} are calculated and the results are stored in the database. The data is publicly available and easy to browse in a web-application as will be described in Sec. \ref{subsec:webpanel}.\par 
We note, that selected systems have been subject to excited state calculations in order to obtain ZPL energies, PL spectra, HR factors, \textit{etc.} enabling the identification of promising defect candidates for optical applications as is discussed in Sec. \ref{subsec:quantum_applications}.

\subsection{The QPOD database}
The QPOD database uses the ASE DB format \cite{larsen2017atomic} which currently has five backends: \texttt{JSON}, \texttt{SQLite3}, \texttt{PostgreSQL}, \texttt{MySQL}, \texttt{MariaDB}. An ASE DB enables simple querying of the data \textit{via} the in-built \texttt{ase db} command line tool, a Python interface, or a webapp (see Sec. \ref{subsec:webpanel}). With those different possibilities to access and interact with the data, we aim to give users a large flexibility based on their respective technical background and preferences.\par 
Each row of QPOD is uniquely defined by defect name, host name, and it's respective charge state. Furthermore, the fully relaxed structure as well as all of the data associated with the respective defect is attached in the form of key-value pairs or \texttt{JSON}-formatted raw data.
\subsection{The QPOD webapp}\label{subsec:webpanel}
A defining feature of the QPOD database is its easy accessibility through a web application (webapp). For each row of the database, one can browse a collection of web-panels designed to highlight the various computed properties of the specific defect. Specific elements of the web-panels feature clickable "?" icons with explanatory descriptions of the content to improve the accessibility of the data.\par 
Directing between different entries of the database is either possible by using hyperlinks between related entries, or using the overview page of the database, where the user can search for materials, and order them based on different criteria. Furthermore, we ensure the direct connection to C2DB with hyperlinks between a defect material and its respective host material counterpart in C2DB in case users want to find more information about the defect-free systems.
\subsection{Overview of host materials}\label{subsec:hosts}
\begin{figure}
  \centering
  \includegraphics{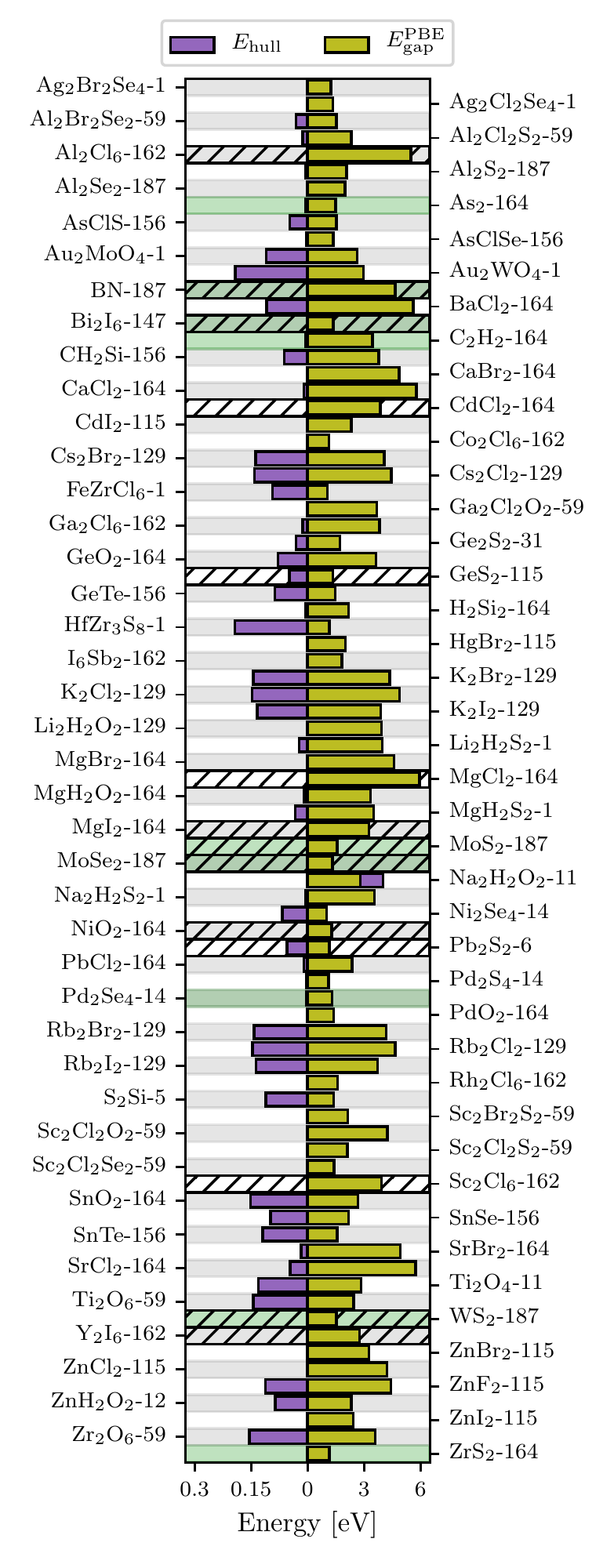}
  \caption{\textbf{Overview of host crystals.} Energy above convex hull in eV/atom $E_\mathrm{hull}$ (purple bars, left) and PBE-calculated band gap $E_\mathrm{gap}^\mathrm{PBE}$ (yellow bars, right) of the $82$ host crystals of QPOD. Host materials which have been realized experimentally (in their monolayer form) are highlighted in green and materials with a known layered bulk phase and related ICSD \cite{bergerhoff1983inorganic,levin2018nist} entry are shown with a hatch pattern.}
  \label{fig:3.4_hosts}
\end{figure}
In total, $82$ host crystals that were chosen according to the criteria outlined in Sec. \ref{subsec:workflow_QPOD} comprise the basis of our systematic study of intrinsic point defects. The set of host materials span a range of crystal symmetries, stoichiometries, chemical elements, and and band gaps (see Fig. \ref{fig:3.4_hosts}). Among them, at least nine have already been experimentally realized in their monolayer form, namely As$_2$, BN, Bi$_2$I$_6$, C$_2$H$_2$, MoS$_2$, MoSe$_2$, Pd$_2$Se$_4$, WS$_2$, and ZrS$_2$ whereas $15$ possess an ICSD \cite{bergerhoff1983inorganic,levin2018nist} entry and are known as layered bulk crystals. The PBE band gaps range from $1.02$ eV for Ni$_2$Se$_4$ up to $5.94$ eV for MgCl$_2$ making our set of starting host crystals particularly heterogeneous.

\section{Results}\label{sec:results}
In this section we first present some general illustrations and analyses of the data in QPOD. We then leverage the data to address three specific scientific problems, namely the identification of: (i) Defect tolerant semiconductors with low concentrations of mid-gap states. (ii) Intrinsically \textit{p}-type or \textit{n}-type semiconductors. (iii) Optically accessible high-spin defects for quantum technological applications.    
\subsection{Relaxation of defect structures}
A major part of the computational efforts to create the QPOD went to the relaxation of the defect structures in a symmetry broken supercell. As discussed in Section \ref{subsec:supercell} the strategy to actively break the symmetry of the host crystal by the choice of supercell was adopted to enable defects to relax into their lowest energy configuration.
\par
\begin{figure*}
  \centering
  \includegraphics{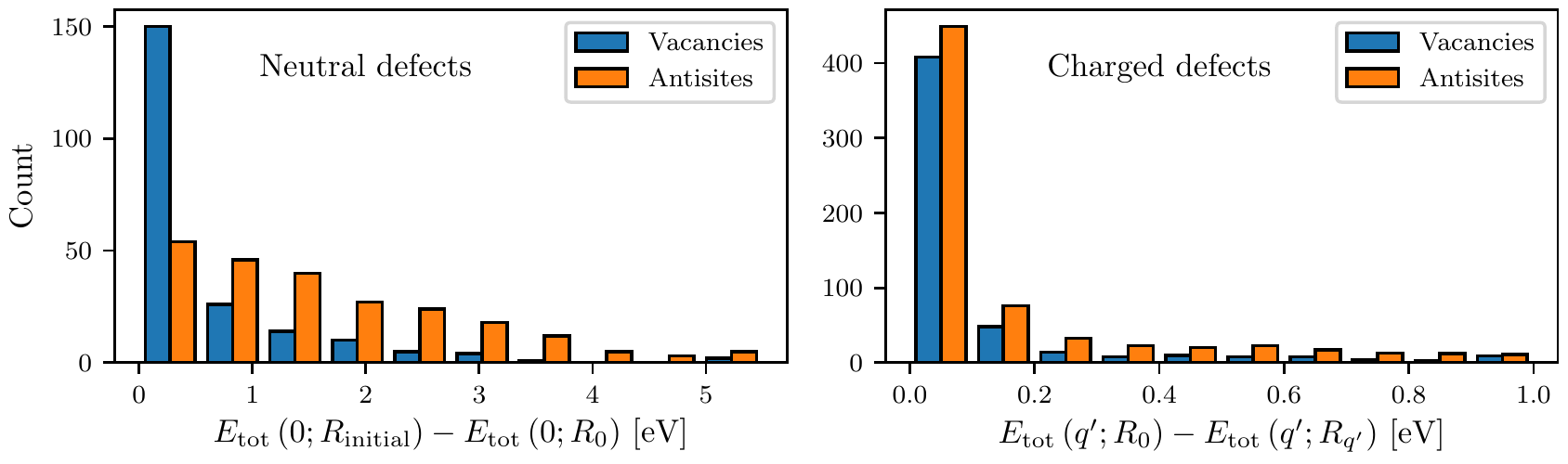}
  \caption{\textbf{Relaxation effects for the formation of charged and neutral defects.}  Left: histogram of the reorganization energy from the initial defect substitution to the neutral equilibrium configuration. Low values on the $x$-axis correspond to small reorganization of defect structures upon addition of a defect to the pristine host crystal. Right: histogram of the reorganization energy between a neutral defect and its charged counterpart.}
  \label{fig:4.1_relaxation}
\end{figure*}

Figure \ref{fig:4.1_relaxation} shows the gain in total energy due to the relaxation for the over $1900$ vacancy and antisite defects (different charge states included). Not unexpectedly, the relaxation has the largest influence on antisite defects while vacancy structures in general show very weak reorganization relative to the pristine structure as can be seen in the left panel of Fig. \ref{fig:4.1_relaxation}. The relaxations for charged defects have always been started from the neutral equilibrium configuration of the respective defect. As a result, the reorganization energies for charged defects is significantly lower (see right panel of Fig. \ref{fig:4.1_relaxation}).

\subsection{Charge transition levels}
\begin{figure*}
  \centering
  \includegraphics{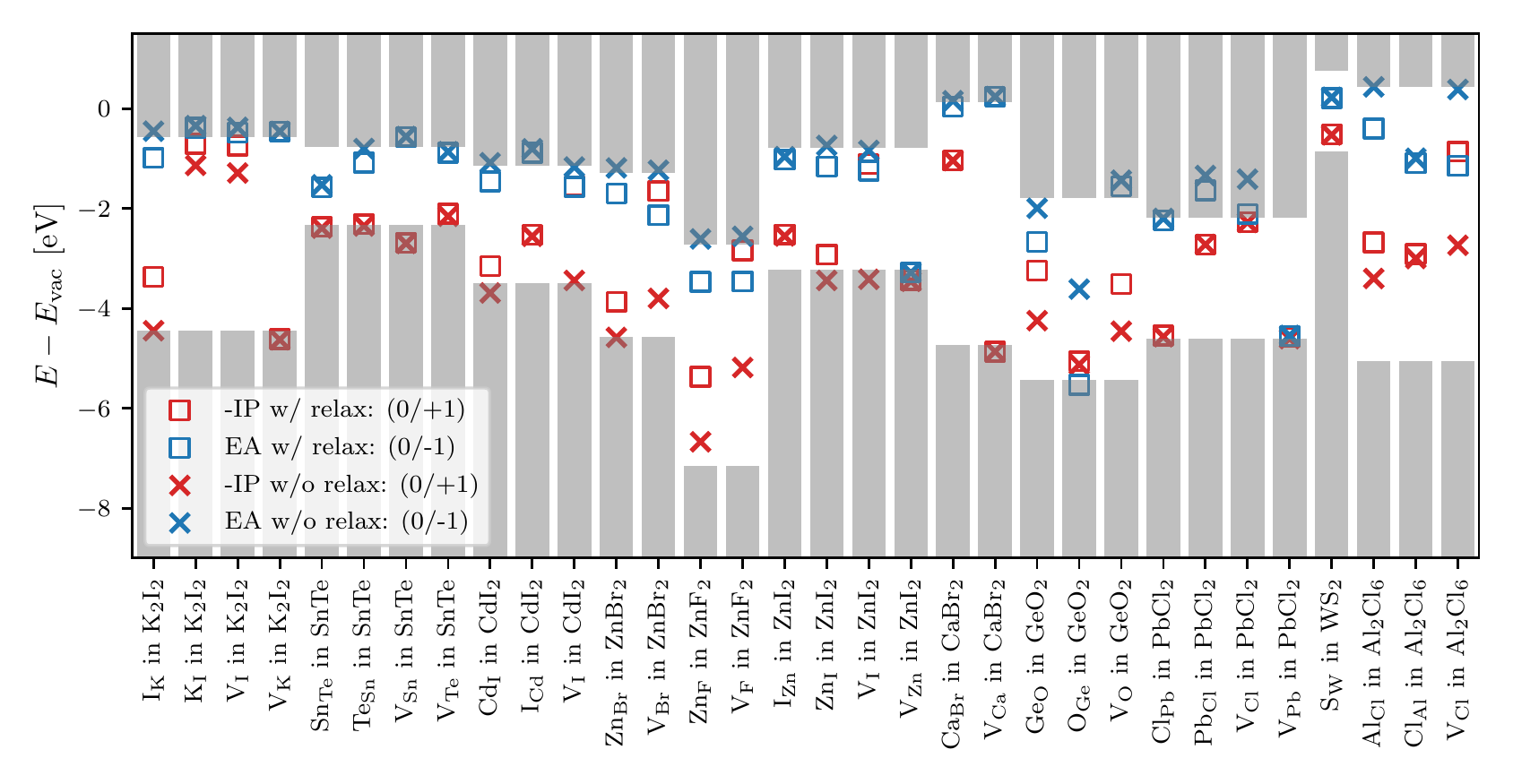}
  \caption{\textbf{Relaxation effects for ionisation potentials and electron affinities.} Energies of -IP (red symbols) and EA (blue symbols) with and without relaxation effects included (boxes and crosses, respectively). The energies are all referenced to the vacuum level of the pristine crystal and grey bars represent the valence/conduction band of the individual host crystals.}
  \label{fig:4.2_ipea}
\end{figure*}
In Section \ref{sec:SJ} we described how we obtain the CTLs by combining Slater-Janak transition state theory on a static lattice with geometry relaxations in the final state. For "negative" transitions, e.g. $\varepsilon\left(0/-1\right)$, the effect of the relaxation is to lower the energy cost of adding the electron, \textit{i.e.} the reorganisation energy lowers the CTL. In contrast, the lattice relaxations should produce an upward shift for "positive" transitions, \textit{e.g.} $\varepsilon\left(0/+1\right)$, because in this case the CTL denotes the negative of the energy cost of removing the electron.  
\par
Figure \ref{fig:4.2_ipea} shows the $\varepsilon\left(0/+1\right)$ and $\varepsilon\left(0/-1\right)$ CTLs for a small subset of defects. Since the energies are plotted relative to the vacuum level, the CTLs correspond to the (negative) ionisation potential (-IP) and electron affinity (EA), respectively. Results are shown both with and without the inclusion of relaxation effects. As expected, the relaxation always lowers the EA and the IP (the $\varepsilon\left(0/+1\right)$ is always raised). The reorganisation energies can vary from essentially zero to more than 2 eV, and are absolutely crucial for a correct prediction of CTLs and (charged) defect formation energies.  

We notice that the CTLs always fall inside the band gap of the pristine host (marked by the grey bars) or very close to the band edges. This is clearly expected on physical grounds, as even for a defective system the CTLs cannot exceed the band edges (there are always electrons/holes available at the VBM/CBM sufficiently far away from the point defect). However, for small supercells such behavior is not guaranteed as the band gap of the defective crystal could deviate from that of the pristine host material. Thus, the fact that the CTLs rarely appear outside the band gap is an indication that the employed supercells are generally large enough to represent an isolated defect.  

When the Fermi level is moved from the VBM to the CBM one expects to fill available defects states with electrons in a stepwise manner, \textit{i.e.} such that $\varepsilon\left(q+1/q\right) < \varepsilon\left(q/q-1\right)$. In particular, we expect $-\mathrm{IP}<\mathrm{EA}$.  Interestingly, for a few systems, \textit{e.g.} V\textsubscript{F} in ZF$_2$, the ordering of IP and EA is inverted. The physical interpretation of such an ordering is that the neutral charge state becomes thermodynamically unstable with respect to positive and negative charge states \cite{freysoldt2014first}. This results in a direct transition from positive to negative charge state in the formation energy diagram.

\subsection{Intrinsic carrier concentrations}\label{sec:dopability}
For many of the potential applications of 2D semiconductors, \textit{e.g.} transistors\cite{radisavljevic2011single}, light emitting devices\cite{withers2015light}, or photo detectors\cite{koppens2014photodetectors}, the question of dopability of the semiconductor material is crucial. Modulation of the charge carrier concentration is a highly effective means of controlling the electrical and optical properties of a semiconductor. This holds in particular for 2D semiconductors whose carrier concentrations can be modulated in a variety of ways including electrostatic or ionic gating\cite{wang2012electronics,wang2015ionic}, ion intercalation\cite{yue2021ionic}, or surface functionalisation\cite{xiang2015surface}. In general, these methods are only effective if the material is not too heavily doped by its ubiquitous native defects, which may pin the Fermi level close to one of the band edges. For applications relying on high carrier conduction rather than carrier control, a high intrinsic carrier concentration may be advantageous - at least if the native defects do not degrade the carrier mobility too much, see Sec. \ref{sec:defecttolerance}.
\par
\begin{figure}
  \centering
  \includegraphics{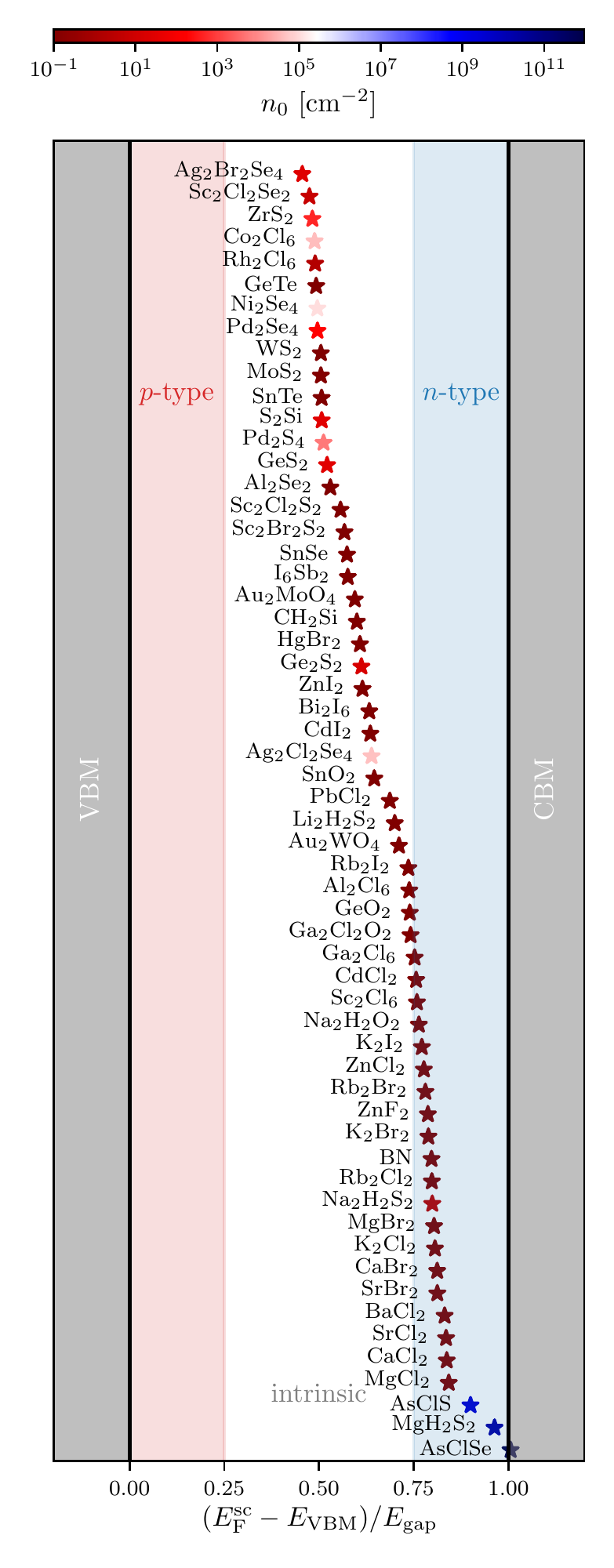}
  \caption{\textbf{Self-consistent Fermi-level position for host materials within the pristine band gap.} The energy scale is normalized with respect to each individual pristine band gap. The regions close to the CBM, VBM  correspond to the \textit{p}-, \textit{n}-type dopable regimes, respectively. The colorcode of the markers represents the equilibrium electron carrier concentration $n_0$.}
  \label{fig:4.3_dopability}
\end{figure}

Figure \ref{fig:4.3_dopability} shows the calculated position of the equilibrium Fermi level (at room temperature) for all the 2D materials considered as defect hosts in this work. It should be noted that the Fermi level position depends on the number of different defect types included in the analysis. Consequently, the results are sensitive to the existence of other types of intrinsic defects with formation energies lower than or comparable to the vacancy and antisite defects considered here. Fermi level regions close to the VBM or CBM  (indicated by red/blue colors) correspond to $p-$type and $n-$type behavior, respectively, while Fermi levels in the central region of the band gap correspond to intrinsic behavior. 

Clearly, most of the materials present in the QPOD database show either intrinsic or \textit{n}-type behavior. An example of a well known material with intrinsic behavior is MoS$_2$, where native defects pin the Fermi level deep within the band gap resulting in very low electron and hole carrier concentrations in good agreement with previous observations\cite{shang2018elimination,splendiani2010emerging,noh2014stability,komsa2015native}. As an example of a natural \textit{n}-type semiconductor we highlight the Janus monolayer AsClSe, which presents an impressive electron carrier concentration of $9.5 \times 10^{11}$ cm$^{-2}$ at 300 K, making it an interesting candidate for a high-conductivity 2D material. For the majority of the materials, the equilibrium carrier concentrations are in fact relatively low implying a high degree of dopability. We note that none of the materials exhibit intrinsic $p$-type behavior. This observation indicates that the challenge of finding naturally $p$-doped semiconductors/insulators, which is well known for bulk materials\cite{fortunato2012oxide,kormath2018database,xu2018prediction,bhatia2016high}, carries over to the class of 2D materials.

\subsection{Defect formation energies: Trends and correlations}
\begin{figure}
  \centering
  \includegraphics{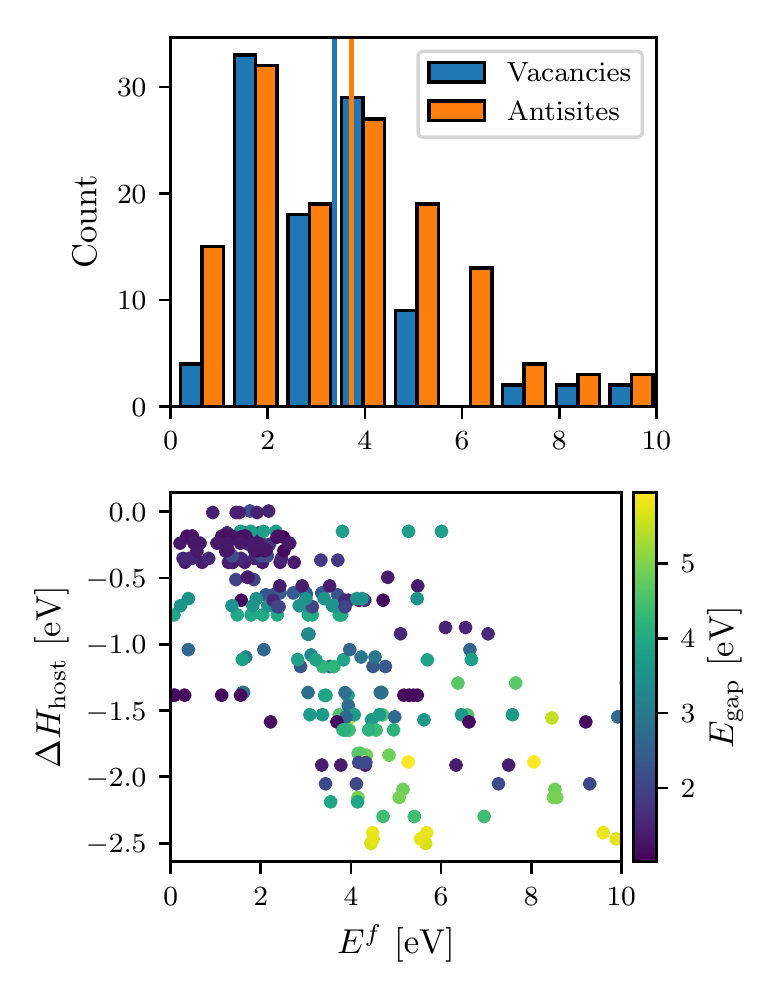}
  \caption{\textbf{Distribution of defect formation energies in the neutral charge state.} Top: histogram of neutral formation energies $E^f$ wrt. standard states for vacancy and antisite defects. The vertical lines represent the mean value. Bottom: heat of formation $\Delta H_\mathrm{host}$ of the pristine monolayer as a function of the neutral formation energies with the pristine PBE-calculated band gap as a color code.}
  \label{fig:4.4_formation}
\end{figure}
The formation energy is the most basic property of a point defect. Figure \ref{fig:4.4_formation} the distribution of the calculated formation energies of neutral defects for both vacancy (blue) and antisite (orange) defects. For the chemical potential appearing in Eq. (\ref{eq:eform}) we used the standard state of each element. There is essentially no difference between the two distributions, and both means (vertical lines) are very close to 4 eV. Roughly half ($44$ \%) of the neutral point defects have a formation energy below $3$ eV and $28$ \% are below $2$ eV. This implies that many of the defects would form readily during growth and underlines the importance of including intrinsic defects in the characterization of 2D materials. As a reference, the NV center in diamond shows formation energies on the order of $5$ eV to $6$ eV\cite{deak2014formation} (HSE value).

The lower panel of Figure \ref{fig:4.4_formation} shows the defect formation energy relative to the heat of formation of the pristine host material, $\Delta H_{\mathrm{host}}$. There is a clear correlation between the two quantities, which may not come as a surprise since the $\Delta H_{\mathrm{host}}$ measures the gain in energy upon forming the material from atoms in the standard states. 
More stable materials, \textit{i.e.} materials with more negative heat of formation, are thus less prone to defect formation than less stable materials. It is also interesting to note the correlation with the band gap of the host material, indicated by the color of the symbols. A large band gap is seen to correlate with a large (negative) $\Delta H_{\mathrm{host}}$ and large defect formation energies, and \textit{vice versa} a small band gap is indicative of a smaller (negative) $\Delta H_{\mathrm{host}}$ and low defect formation energies. These trends are somewhat problematic as low band gap materials with long carrier lifetimes, and thus low defect concentrations, are required for many applications in (opto)electronics, while large band gap host materials with high density of (specific types of) defects are required for many color center-based quantum technology applications.


\subsection{Point defect symmetries}
\begin{figure}
  \centering
  \includegraphics{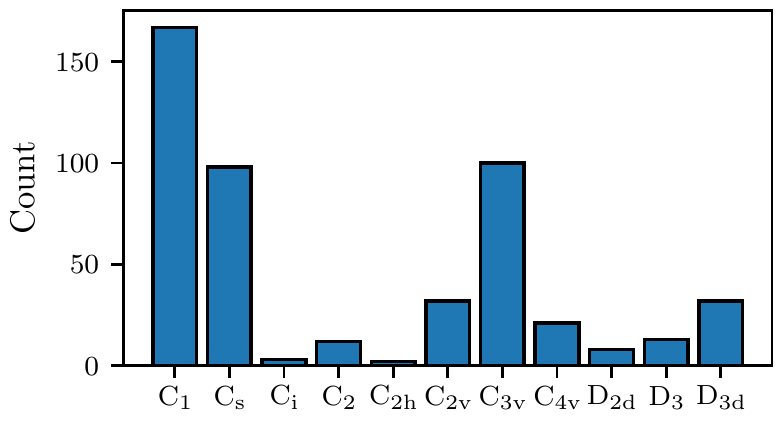}
  \caption{\textbf{Distribution of point groups for relaxed defects.} Point groups are ordered from the lowest symmetry group (i.e. C$_\mathrm{1}$) to the highest occurring symmetry group (D$_{3d}$).}
  \label{fig:4.5_symmetry}
\end{figure}
Another basic property of a point defects is its local symmetry, which determines the possible degeneracies of its in-gap electronic states and defines the selection rules for optical transitions between them. Figure \ref{fig:4.5_symmetry} shows the distribution of point groups (in Sch\"onflies notation) for all the investigated neutral defect systems. A large fraction of the defects break all the symmetries of the host crystal ($34$ \% in C$_1$) or leave the system with only a mirror symmetry ($20$ \% in C$s$). A non-negligible number ($31$ \%) of defects can be characterized by $2$, $3$, or $4$-fold rotation axis with vertical mirror planes (C$_{2v}$, C$_{3v}$ and C$_{4v}$). Naturally, C$_{3v}$ defects ($20$ \% overall) often stem from hexagonal host structures, some examples being V\textsubscript{S} and W\textsubscript{S} in WS$_2$, V\textsubscript{Se} in SnSe, and Ca\textsubscript{Br} in CaBr$_2$. Such defects share the symmetry group of the well known NV center in diamond\cite{lenef1996electronic} and might be particularly interesting candidates for quantum technology applications. Relatively few defect systems ($3$ \%) incorporate perpendicular rotation axes, \textit{e.g.} the D$_3$ symmetry of I\textsubscript{Sb} in I$_6$Sb$_2$.\par 
Adding or removing charge to a particular defect system can influence the structural symmetry as it was previously observed for the negatively charged sulfur vacancy in MoS$_2$\cite{tan2020stability}. We find that $10$ \% of the defects in QPOD undergo a change in point group when adding (removing) an electron to (from) the neutral structure and relaxing it in its respective charge state.

\subsection{Defect tolerant materials}\label{sec:defecttolerance}
Although defects can have useful functions and applications, they are often unwanted as they tend to deteriorate the ideal properties of the perfect crystal. Consequently, finding defect tolerant semiconductors\cite{pandey2016defect,walsh2017instilling}, \textit{i.e.} semiconductors whose electronic and optical properties are only little influenced by the presence of their native defects, is of great interest. 

When discussing defect tolerance of semiconductors one should distinguish between two different situations: (i) For transport applications where the system is close to equilibrium, defects act as scattering centres limiting the carrier mobility. In this case, charged defects represent the main problem due to their long range Coulomb potential, which leads to large scattering cross sections. (ii) For opto-electronic applications where relying on photo-excited electron-hole pairs, deep defect levels in the band gap represent the main issue as they facilitate carrier capture and promote non-radiative recombination. In the following we examine our set of host materials with respect to type (ii) defect tolerance.\par
\begin{figure}
  \centering
  \includegraphics{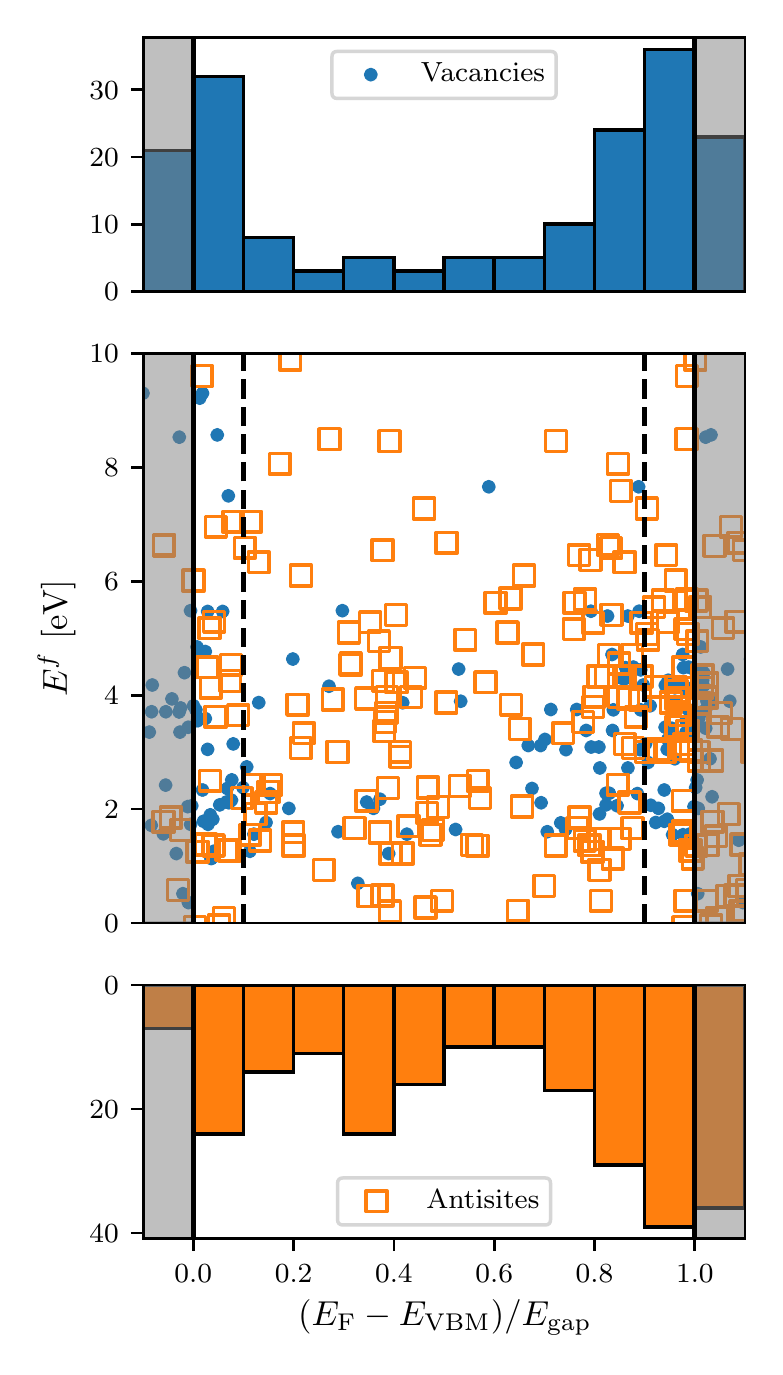}
  \caption{\textbf{Defect tolerances for vacancy and antisite defects.} Position of defect charge transition levels as a function of Fermi energy (normalized with respect to the band gap) for antisite defects (orange squares) and vacancies (blue dots). Top (bottom): histogram of the occurance of a CTL for vacancies (antisites) within a certain energy range. If a defect's CTL lies above $0.9$/below $0.1$ (dashed black vertical lines in the middle plot) it corresponds to a CTL not having a detrimental effect on the host's properties. Materials with these CTL are considered defect-tolerant wrt. optical properties. The grey areas on the left and right hand side represent the valence band and conduction band, respectively.}
  \label{fig:4.6_tolerances}
\end{figure}

 
Figure \ref{fig:4.6_tolerances} shows the positions of charge transition levels of vacancy (blue) and antisite (orange) defects as a function of the Fermi level normalized to a host material's band gap. The neutral formation energy of the defects is shown in the middle panel, where we have also marked regions of shallow defect states that lie within 10\% of the band edges (black dashed lines). A host material is said to be type (ii) defect tolerant if all of its intrinsic defects are shallow or all its deep defects have high formation energy. A number of defect tolerant host materials are revealed by this analysis, including the ionic halides K$_2$Cl$_2$, Rb$_2$I$_2$, and Rb$_2$Cl$_2$. With formation energies lying about 150 meV/atom above those of their cubic bulk structures, these materials may be challenging to realise in atomically thin form. Nevertheless, their defect tolerant nature fits well with the picture of deep defect states having larger tendency to form in covalently bonded insulators with bonding/anti-bonding band gap types compared to ionic insulators with charge-transfer type band gaps\cite{pandey2016defect}.
\par
The data in Figure \ref{fig:4.6_tolerances} suggests that vacancy and antisite defects have different tendencies to form shallow and deep defect states, respectively: While $55$ \% of the vacancy defects form shallow defect states, this only happens for $30$ \% of the antisites. The trend can be seen in the top and bottom histograms of Fig. \ref{fig:4.6_tolerances} where the vacancy distribution (top) shows fewer CTLs around the center of the band gap compared to the antisite distribution (bottom). Based on this we conclude that vacancy defects are, on average, less detrimental to the optical properties of semiconductors than antisite defects.  
\par

\subsection{Defects for quantum technological applications}\label{subsec:quantum_applications}
Point defects with a high-spin ground state, \textit{e.g.} triplets, are widely sought-after as such systems can be spin initialized, manipulated and read out (the N$_\mathrm{V}^-$ centre in diamond is a classical example). In particular, triplet spin systems may be exploited in optically detected magnetic resonance spectroscopy to act as qubits, quantum magnetometers, and other quantum technological applications\cite{eckstein2013materials,gardas2018defects}. 

\begin{figure}
  \centering
  \includegraphics{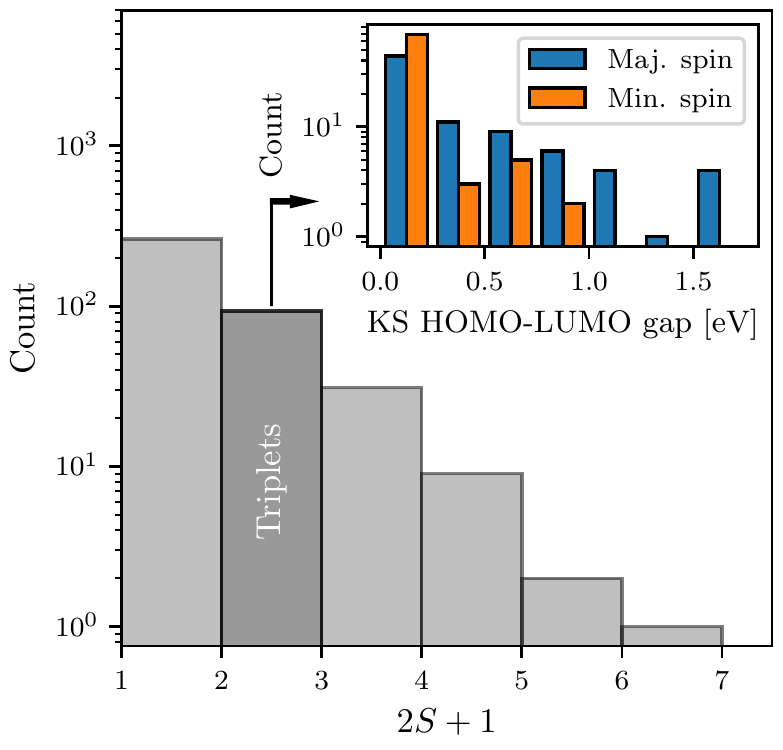}
  \caption{\textbf{An overview of distribution of total magnetic moments for all defect systems.} The bin corresponding to systems with triplet ground state is highlighted. A finite Fermi-smearing in our calculations gives rise to spin-contamination and hence the absolute number of entries in each bin is somewhat inaccurate. The inset shows the distribution of Kohn-Sham HOMO-LUMO gaps in both spin-channels for the triplets. The systems with a Kohn-Sham HOMO-LUMO gap of at least $0.5$ eV are chosen for excited states calculations.}
  \label{fig:magnetic}
\end{figure}

An overview of the distribution of local magnetic moments for all the defect systems is shown in Fig. \ref{fig:magnetic}. Although defects with any spin multiplicity are of interest and have potential applications, \textit{e.g.} as sources of intense and bright single photon emission, we have limited our detailed excited state investigations to the 80 defects with a triplet ground state. These systems are subsequently screened for a (PBE) Kohn-Sham gap of at least 0.5 eV.  After this screening, we are left with 33 systems. Out of these 33 defect systems, we further discard those involving heavy atoms, \textit{e.g.} PbCl$_2$, because of large spin-orbit coupling effects that would hinder their usage in quantum technological applications, leaving us with a total of 25 systems, which are subject to excited state calculations (see Sec. \ref{subsec:excited_methodology}). The excited state calculations are performed with the motivation of finding systems exhibiting sharp radiative transitions in the mid infrared to ultraviolet frequency range. 

 Most of these 25 defects exhibit large structural changes upon optical excitation. After selecting the defects with the smallest $\Delta Q$ values, we are left with 4 systems for which we calculate the PL line shape. A detailed study of these 4 systems, \textit{e.g.} their optical cycles, inter-system crossing rates, spin-coherence times, will be the subject of a future study. In the present work we limit ourselves to the calculation of their PL line shape, which is a key characteristic of point defects, which may be used to identify their precise atomic structure\cite{sajid2020vncb}.

\begin{figure}
  \centering
  \includegraphics{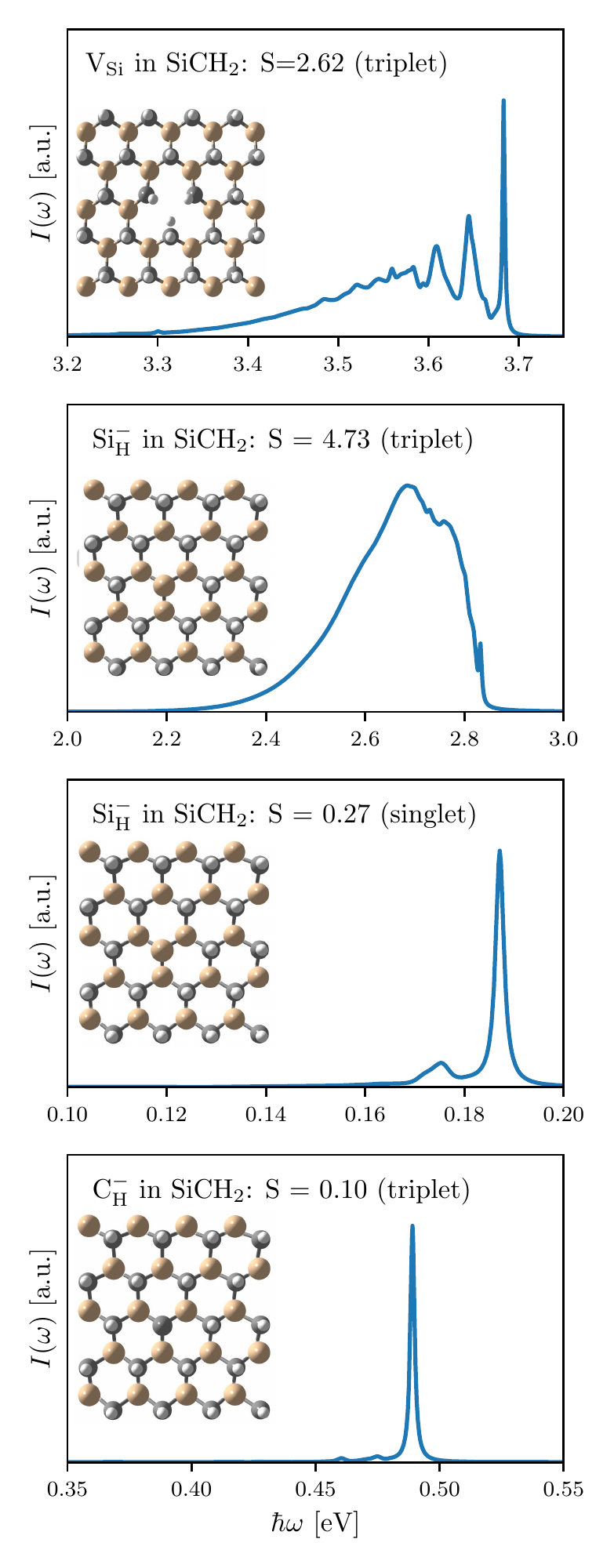}
  \caption{\textbf{PL lineshapes for a few selected systems using generating function approach.} A few selected systems with small Huang-Rhys factors are shown. The top panel show the PL line shape for HOMO to LUMO transition in the majority spin channel for V$_\mathrm{Si}$ defect in SiCH$_{2}$. The middle two panels shows the PL spectra for HOMO to LUMO transition in the minority spin channel (2nd panel) and HOMO to LUMO transition within the singlet manifold (3rd panel)  for Si$_\mathrm{H}^{-}$ defect in SiCH$_2$ host material. The bottom panel shows the calculated spectra for C$_\mathrm{H}^{-}$ defect in SiCH$_{2}$ for HOMO to LUMO transition in the majority spin channel. The calculated radiative life times of these transitions (top to bottom) are $1.3\times10^{-3}$ ns, $4.9\times10^{-2}$ ns, $2.9\times10^{-4}$ ns and $9.0\times10^{-4}$ ns, repectively.}
  \label{fig:4.8_lineshapes}
\end{figure}

The PL lineshapes calculated using the generating function approach are shown Figure \ref{fig:4.8_lineshapes} for the lowest triplet transitions of V$_\mathrm{Si}$, C$_\mathrm{H}^-$ and Si$_\mathrm{H}^-$ in SiCH$_2$ and the lowest singlet transition of Si$_\mathrm{H}^-$ in SiCH$_2$. The upper and bottom panels show the line shape for the HOMO-LUMO transition in the majority spin channel for V$_\mathrm{Si}$ and C$_\mathrm{H}^{-}$, respectively. While the Huang-Rhys factors for V$_\mathrm{Si}$ is small, it is extremely small for C$_\mathrm{H}^{-}$. In fact, radiative transitions with such small line widths have rarely been reported previously in 2D semi-conductors\cite{sajid2020single}, which makes these results promising. The second panel shows the PL spectra for the lowest transition in the minority spin channel of Si$_\mathrm{H}^{-}$, while the third panel shows the PL spectrum for the lowest transition within the singlet manifold of the same defect, calculated using Eq. (\ref{eq:OSS}). The symmetry of this defect, \textit{i.e.} $C_{3v}$, and the KS level structure is somewhat similar to the N$_\mathrm{V}^-$ centre in diamond. The fact that the calculated Huang-Rhys factor is also close to the one reported for N$_\mathrm{V}^-$ \cite{alkauskas2014first} makes this defect system particularly interesting as a 2D counter part of the highly useful N$_\mathrm{V}^-$ centre.  

\par

\section{Summary and outlook}\label{sec:summaryoutlook}
In summary, more than $1900$ intrinsic point defects hosted by $82$ different 2D semiconductors and insulators have been relaxed and characterised by means of DFT calculations. The $1900+$ point defects comprise about 500 structurally distinct vacancy and antisite defects in different charge states. The thousands of DFT calculations were orchestrated by a computational workflow built using the Atomic Simulation Recipes (ASR)\cite{gjerding2021atomic} and executed with the MyQueue\cite{mortensen2020myqueue} task scheduler.

The ASR defect workflow includes the calculation of formation energies, charge transition levels, equilibrium defect and carrier concentrations, point group symmetry labels of in-gap states, transition dipole moments, hyperfine coupling, and zero field splittings. All DFT calculations were performed using the semi-local PBE xc-functional. While the PBE should yield accurate structural and thermodynamic properties, its tendency to underestimate band gaps can result in high-energy defect levels being missed, if they lie above the PBE band gap. We emphasise, however, that charge transition levels appearing within the PBE band gap should be well described by our PBE-based Slater-Janak transition state approach.  

The analysis of defect formation energies showed that many of the investigated defects are likely to be present in the host material in appreciable concentrations thus confirming their relevance and importance for the general properties of the host materials. Based on the thermodynamic and electronic characterisation of all the intrinsic defects, we identified a number of defect tolerant ionic insulators among the host materials. While these specific materials may be challenging to synthesize due to competing bulk phases, the results can be used to guide future searches for defect tolerant semiconductors for high-performance opto-electronic 2D devices. Among the 82 host materials, we found several semiconductors with high intrinsic electron carrier concentrations, \textit{e.g.} MgS$_2$H$_2$ and the Janus materials AgClS and AgClSe, whereas no intrinsic \textit{p}-type materials were found.

Out of the $>1900$ defects, only around 80 adopt a high-spin ($S>1/2$) ground state and only a few defects showed low Huang-Rhys factors and correspondingly narrow photoluminescence spectra making them relevant as single-photon emitters. This indicates that the simple defect types considered in this work are not likely to yield useful spin defects for quantum technology applications. In this work, the excited state properties were calculated and analyzed manually. Incorporating this part into the automated workflow will be an important future extension of the current methodology that is critical to enable a systematic and rational design of defects with ideal excited state properties including transition energies, excited state lifetimes/dynamics, and emission line shapes.

The web-based presentation of the QPOD database and its seamless integration with the Computational 2D Materials Database (C2DB) makes a unique platform for exploring the physics of point defects in 2D materials, which should be useful both as a convenient lookup table and as a benchmark reference for computational studies. Moreover, the possibility to download the entire database makes it applicable for machine learning purposes, which has a large yet untapped potential for establishing structure-property relationships for point defects. 

Looking ahead, there are many possible extensions and improvements of the current work. First of all, it would be interesting to move beyond the PBE to more advanced xc-functionals, such as screened hybrids. This would not only enhance the data quality/accuracy but also provide the basis for a systematic and statistically significant assessment of the performance of the PBE for defect calculations. It would also be relevant to expand the set of host materials beyond the 82 materials considered here. The C2DB currently contains about 500 monolayers exfoliated from experimentally known layered van der Waals crystals and a similar number of predicted highly stable monolayers providing ample opportunities for selecting ideal 2D host crystals. As mentioned, the simple defects considered in this work turned out to be mostly non-magnetic. Thus, for applications relying on spinful defect centers, \textit{e.g.} magnetic field sensors or qubits, it seems important to incorporate more complex defects such as divacancies and vacancy-substitutional defect pairs.  

\section{Methods}

\subsection{Density functional theory calculations}
All the DFT calculations (spin-polarized) are performed by the GPAW electronic structure code \cite{enkovaara2010electronic} using a plane wave basis set with $800$ eV plane wave cut off, $k$-point density of $6$ \AA$^{-1}$ ($12$ \AA$^{-1}$) for structural relaxations (for ground state calculations) and the PBE xc-functional \cite{perdew1996generalized}. The supercell is kept fixed and atoms are fully relaxed until forces are below $0.01$ eV/\AA. We apply a Fermi smearing of $0.02$ eV ($0.2$ eV for relaxations) for all systems and increase that parameter slightly ($0.05$, $0.08$ or $0.1$) for systems whose ground state proves difficult to converge. We use the Pulay mixing scheme \cite{pulay1980convergence} where total density and magnetisation densities are treated separately. The excited states are calculated, at gamma point, using the same computational parameters as the ground state, and using the DO-MOM method\cite{levi2020variational}, where the maximum step length, $\rho_\mathrm{max}$ , for the quasi-newton search direction is chosen to be $0.2$. The parameters for the Slater-Janak calculations are the same as for the ground state calculations.\par 
An initial benchmarking (for formation energies, CTLS, HF, ZFS, TDM, excited states, \textit{etc.})was performed on a subset of defect systems prior to the work highlighted here, in order to ensure that our methods, implementations and calculations yield reasonable results.

\section{Data availability}
The QPOD database will be available upon request. The web-application of the database will be publicly available when the paper is published.

\section{Code availability}
The ASR recipe scripts used in the QPOD workflow are available at: \url{https://gitlab.com/asr-dev/asr/-/tree/defect\_formation/asr}. The scripts to generate the figures in this paper are available at: \url{https://gitlab.com/fBert31/defect-paper}.

\section*{References}
\bibliographystyle{iopart-num}
\bibliography{references}

\section{Acknowledgments}
The Center for Nanostructured Graphene (CNG) is sponsored by The Danish National Research Foundation (project DNRF103). We acknowledge funding from the European Research Council (ERC) under the European Union’s Horizon 2020 research and innovation program Grant No. 773122 (LIMA) and Grant agreement No. 951786 (NOMAD CoE). K. S. T. is a Villum Investigator supported by VILLUM FONDEN (grant no. 37789).

\section{Author contributions}
F.B. and K.S.T. developed the initial concept, F.B. developed the workflow and database, F.B., S.A., S.M. conducted initial benchmarks and ran DFT calculations, S.A. and S.M. conducted excited state calculations, S.A. conducted the analysis of magneto-optical properties and the identification of potential defects for quantum technological applications, S.M. conducted the symmetry analysis, F.B. conducted the analysis of relaxation effects, CTLs, dopabilities, defect formation trends, and defect tolerances. K.S.T. supervised the work and helped in interpretation of the results. All authors modified and discussed the paper together.

\section{Competing interests}
The authors declare no competing interests.

\end{document}